\documentclass{aa}

\usepackage{graphicx}
\usepackage{txfonts}
\usepackage{float}
\usepackage{mathtools}
\usepackage{subcaption}
\usepackage[version=4]{mhchem}
\usepackage{gensymb}
\usepackage{tikzsymbols}
\usepackage{rotating}
\usepackage{multirow}
\usepackage{mleftright}

     \def\aap{A\&A}
     \def\apjl{ApJL}
     \def\apjs{ApJS}

\begin{document}

  \title{Understanding the atmospheric properties and chemical composition of the ultra-hot Jupiter HAT-P-7b}

   \subtitle{III. Changing ionisation and the emergence of an ionosphere}

    \titlerunning{The ultra-hot atmosphere of HAT-P-7b: III. }

   \author{Ch. Helling \inst{1,2,3}
         \and
         M. Worters\inst{1,2}
         \and
          D. Samra \inst{1,2}
          \and
     K. Molaverdikhani \inst{4,5}
               \and
          N. Iro
          \inst{6}
                    }

   \institute{Centre for Exoplanet Science, University of St Andrews, North Haugh, St Andrews, KY169SS, UK\\         \email{ch80@st-andrews.ac.uk}
         \and
             SUPA, School of Physics \& Astronomy, University of St Andrews, North Haugh, St Andrews, KY169SS, UK
         \and
         SRON Netherlands Institute for Space Research, Sorbonnelaan 2, 3584 CA Utrecht, NL
        \and
         Max Planck Institute for Astronomy, Königstuhl 17, 69117 Heidelberg, Germany
                  \and 
        Landessternwarte, Zentrum für Astronomie der Universität Heidelberg, Königstuhl 12, 69117 Heidelberg, Germany
    \and
              Institute for Astronomy (IfA), University of Vienna,
              T\"urkenschanzstrasse 17, A-1180 Vienna        }
   \date{Received September 15, 2996; accepted March 16, 2997}


\abstract
{Ultra-hot Jupiters are the hottest close-in exoplanets discovered so far, and present a unique possibility to explore hot and cold chemistry on one object. The tidally locked ultra-hot Jupiter HAT-P-7b has a day/night temperature difference of $\bigtriangleup T \simeq 2500$~K, confining cloud formation to the nightside and efficient ionisation to the dayside. Both have distinct observational signatures.}
{We analyse  plasma and magnetic processes in the atmosphere of the ultra-hot Jupiter HAT-P-7b to investigate the formation of an ionosphere and the possibility of magnetically coupling the atmospheric gas as the base for an extended exosphere. We show which ions and atoms may be used as spectral tracers, and if and where conditions for lightning may occur within the clouds of HAT-P-7b. }
{We use 3D GCM results as input for a kinetic cloud formation code and evaluate characteristic plasma and magnetic coupling parameters, and a LTE radiative transfer is solved for the ionised gas phase.  This study is confined to thermal ionisation only.}
{The ionisation throughout HAT-P-7b's atmosphere varies drastically between day- and nightside. The dayside has high levels of thermal ionisation and long-range electromagnetic interactions dominate over kinetic electron-neutral interactions, suggesting a day-night difference in magnetic coupling. K$^+$, Na$^+$, Li$^+$, Ca$^+$, and Al$^+$ are more abundant than their atomic counterparts on the dayside.   The minimum magnetic flux density for electrons for magnetic coupling is $B<0.5$G for all regions of HAT-P-7b's atmosphere.}
{HAT-P-7b's dayside has an asymmetric ionosphere that extends deep into the atmosphere, the nightside has no thermally driven ionosphere. A corresponding 
asymmetry is imprinted in the ion/neutral composition at the terminators. The ionosphere on HAT-P-7b may be  directly traced by the Ca$^+$ H\&K lines if the local temperature is $\geq$ 5000K. The whole atmosphere may couple to a global, large-scale magnetic field, and lightning may occur on the nightside.}


  \maketitle

\section{Introduction}\label{i}


Ultra-hot Jupiters are a class of the hottest, close-in, giant gas planets discovered thus far (e.g., \citealt{a}; \citealt{b}, \citealt{c}). This group of extrasolar planets  provides a unique window in exoplanet atmospheres as it combines "hot and cold" chemistry on the same planet. 
WASP-18b was one of the first of such ultra-hot Jupiters for which the dominance of heavy ions like Al$^+$, Ti$^+$, and Na$^+$, K$^+$, Ca$^+$   on the dayside  over their atomic counterpart as well as a night/day transition from a H$_2$ to an atomic hydrogen dominated atmosphere was suggested (\citealt{2019A&A...626A.133H}). 
Detection of ionised species in  atmospheres of ultra-hot Jupiters have become possible due to high-resolution spectroscopy with, for example,  CARMENES and HARPS-N in combination with cross-correlation techniques (e.g. \citealt{2019A&A...627A.165H,2020A&A...638A..26S,2020MNRAS.496..504N,2020ApJ...894L..27P,2020ApJ...888L..13T}). From XSHOOTER/VLT data \cite{2017MNRAS.471.1728L} identified the emission lines of Ca$^+$ and Fe$^+$, in addition to other atomic species, from the brown dwarf WD0137-349B, companion to a white dwarf.
 \cite{2014ApJ...796...16K} demonstrate that the hot Jupiter HD\,209458b has a photoionisation-driven dayside ionosphere and that its electron density is higher than in any solar system planetary ionosphere, with a conductivity comparable to the solar chromosphere. 
\cite{2017NatAs...1E.131R} constrained the minimum magnetic field strengths of HAT-P-7b to 6\,G with their MHD simulations and point out that arising Lorentz forces  acting within the atmosphere of HAT-P-7b will disrupt strong eastwards atmospheric winds on the dayside, which reverse and settle into a dayside oscillating pattern with a characteristic time scale of $10^6$s. 

 HAT-P-7b, an ultra-hot Jupiter first discovered in 2008 \citep{discovery}, is  tidally locked which gives rise to an extreme contrast in temperature between the day and night sides,  $\Delta T \simeq 2500$ K (Fig.~\ref{TEMP}).  \cite{hatp7b1} concludes that no cloud particles can form  on the extremely hot ($\geq 2200$K) day side of HAT-P-7b, there is significant cloud formation ongoing on the cooler nightside. 
 In this paper we utilise a theoretical framework presented by \cite{rodriguez2015reference} using a set of fundamental parameters to analyse the ionization and magnetic coupling state of the atmosphere of HAT-P7b. We demonstrate that ultra-hot Jupiters like HAT-P7b may form an ionosphere that reaches deep into the atmosphere and that varies in geometrical extension across the globe. We study the magnetic properties of the atmosphere in order to determine how much of the atmosphere, if any, is coupled to the potential magnetic field.  We further use  the framework of this analysis to make first investigations about the possible lightning occurrence in ultra-hot Jupiters. Lightning is a driver for the Global Electric Circuit on Earth which links different weather areas through global electric circuit system (\citealt{2012SSRv..168..363R,2016SGeo...37..705H,2019JPhCS1322a2028H}).

Section \ref{ii} summarises our approach, including the  plasma parameters analysed. Section \ref{results} presents our results on the possible emergence of an ionosphere and the possibility of electromagnetic interactions within the atmosphere. Section~\ref{iv.iii} discusses  the possibility of lightning. In Sect.~\ref{detectability}, we suggest that the Ca$^+$ H\&K lines are the best candidates to trace a hot ionosphere on HAT-P7b, and that Fe, Al and Na could also be observable.
Section \ref{v} summarises our conclusions on the plasma properties, magnetic properties and the effects of the changing ionisation on HAT-P-7b.

\section{Approach}\label{ii}

This study utilises the hierarchical model approach outlined in  \cite{hatp7b1}, where we apply our kinetic cloud formation model to 97 pre-calculated 1D ($T_{\rm gas}(z), p_{\rm gas}(z), v_{\rm z}(x, y, z)$) profiles extracted from a 3D Global Circulation Model (GCM) solution for HAT-P-7b (Fig.~\ref{TEMP}).  The undepleted element abundances are the solar element abundances, and  cloud formation does deplete (by nucleation and surface growth) or enrich (by evaporation) those elements involved (here: Mg, Si, Ti, O, Fe, Al, Ca, S, C). The resulting gas-phase composition is calculated by applying chemical equilibrium inside these collisional-dominated atmosphere layers (see also Sect. 2.2 in \cite{hatp7b1}).
In the present paper, we use the combined results to evaluate characteristic plasma parameters in order to study the global thermal ionisation and the possibility of an ionosphere, possible magnetic coupling, and address the emergence of lightning on HAT-P-7b. We assume that $T_{\rm e} = T_{\rm gas}$.

\subsection{Ionisation and magnetic coupling}\label{iii}

We apply the frame work presented in \cite{rodriguez2015reference} to analyse the ionization and magnetic coupling state of a planetary atmosphere. A short summary is provided here.

\subsection{Plasma parameter}
For plasma processes to act, the gas needs to be sufficiently ionised.  The simplest measure is the degree of ionisation, $f_{e}$, 
\begin{equation}\label{eq1}
f_{e} = \dfrac{p_{e}}{p_{e}+p_{gas}}.
\end{equation}
 \cite{rodriguez2015reference} propose to use $f_{e} = 10^{-7}$ as the threshold above which a gas  begins to exhibit plasma behaviour in accordance with  laboratory experiments (\citealt{diver2011plasma,fridman2008plasma}). Hence, only a partial ionisation is required for plasma processes to take effect, like those driving non-ideal MHD flows.

The ability of charged particles to oscillate in response to time-varying electric fields in the atmosphere  is studied by calculating the plasma frequency, $\omega_{pe}$,  
\begin{equation}\label{eq2}
\omega_{pe}= \left( \dfrac{n_{e} e^{2}}{\epsilon_{0} m_{e}} \right) ^{1/2}.
\end{equation}
The location  in the atmosphere (for example, in terms of local gas pressure) where electromagnetic interactions begin to dominate over kinetic collisions between neutrals and electrons is determined by $\omega_{\rm pe} \gg \nu_{\rm ne}$, with $\nu_{\rm ne}$ being the 
frequency with which electrons interact with neutrals,
\begin{equation}\label{nune}
\nu_{\rm ne}= \sigma_{\rm gas} n_{\rm gas} \upsilon_{\rm th, e}.
\end{equation}
$\sigma_{\rm gas}$ is the collision cross-section of neutral particles. We assume that HAT-P-7b's atmosphere is hydrogen dominated and we use $\sigma_{\rm gas} = \pi {r_{H_2}}^2$.  We note, however, that parts of the atmosphere are dominated by atomic hydrogen, instead of molecular hydrogen. Hence, the collisional frequency is overestimated by a factor of 4  for example on the dayside of HAT-P-7b in Fig.~\ref{omegaDl}. $n_{gas}$ is the total ambient gas number density  which reflects the change from H$_2$ to H. $\upsilon_{\rm th, e} $ is the electron thermal velocity $(\upsilon_{\rm th, e} = (k_B T_{\rm e}/m_{\rm e})^{1/2})$. 

The distance beyond which the Coulomb force of a charge distribution no longer influences it's surroundings,  the Debye length, $\lambda_{D}$, is the length-scale beyond which a plasma is considered quasi-neutral ($n_{\rm gas} \cong n_{\rm e} \cong n_{\rm i}$ ),
\begin{equation}\label{eq4}
\lambda_{\rm D} = \left( \dfrac{\epsilon_{0} k_{\rm B} T_{\rm e}}{n_{\rm e} e^2} \right) ^{1/2}.
\end{equation}
For length-scales less than the $\lambda_{\rm D}$, a test charge will experience the effects of any charge imbalance inside a Debye sphere. The gas is considered quasi-neutral on scale lengths, L, larger than the Debye length $\lambda_{\rm D}$ (Figure \ref{fig8}), hence electrostatic forces will not influence the gas behaviour. A plasma is quasi-neutral if $\lambda_{\rm D} \ll L$, where L is a characteristic scale length of the plasma.

\begin{figure*}
\begin{tabular}{p{8cm}p{8cm}} \includegraphics[width=1.05\linewidth, page=1]{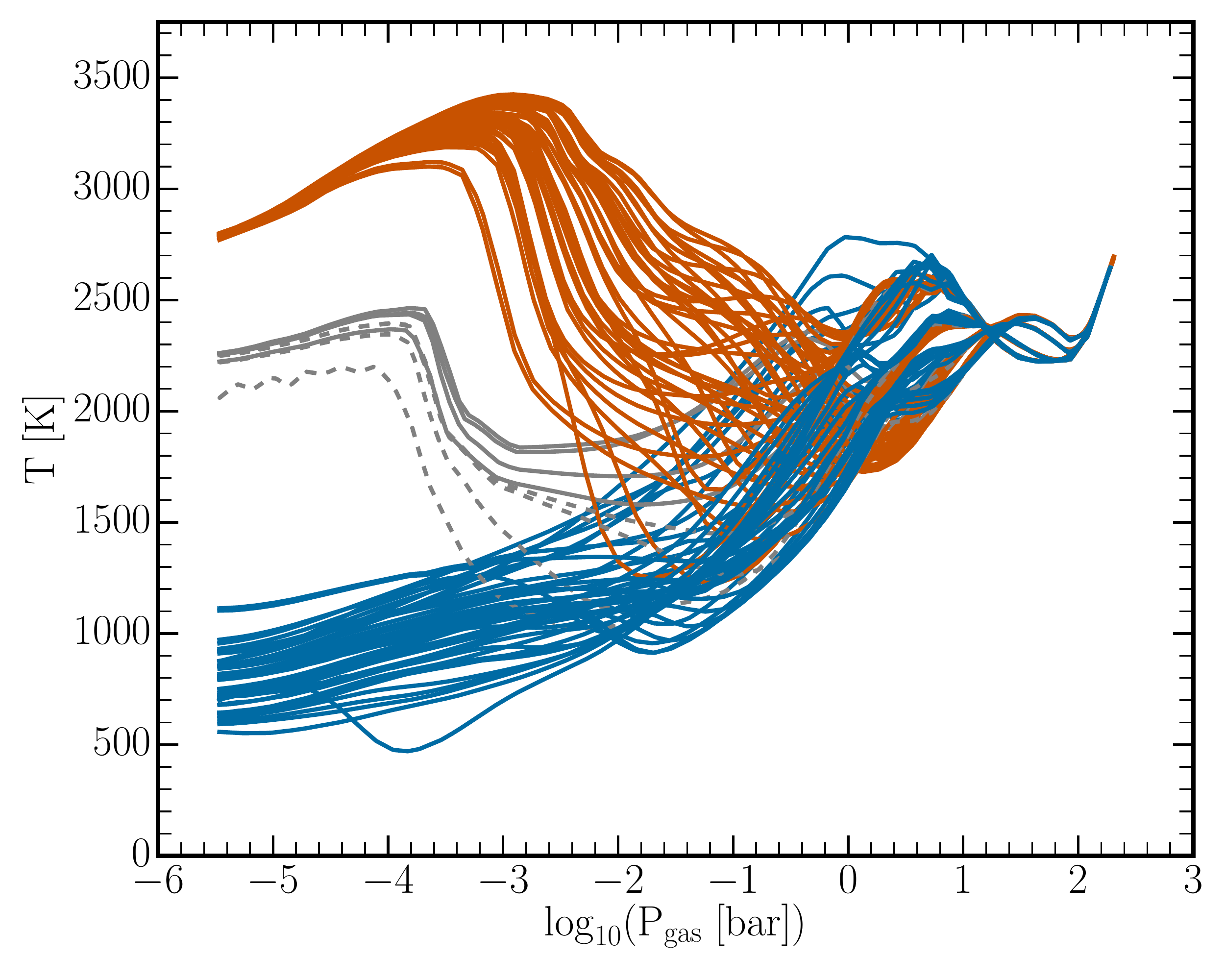} &
  \includegraphics[width=1.05\linewidth,page=1]{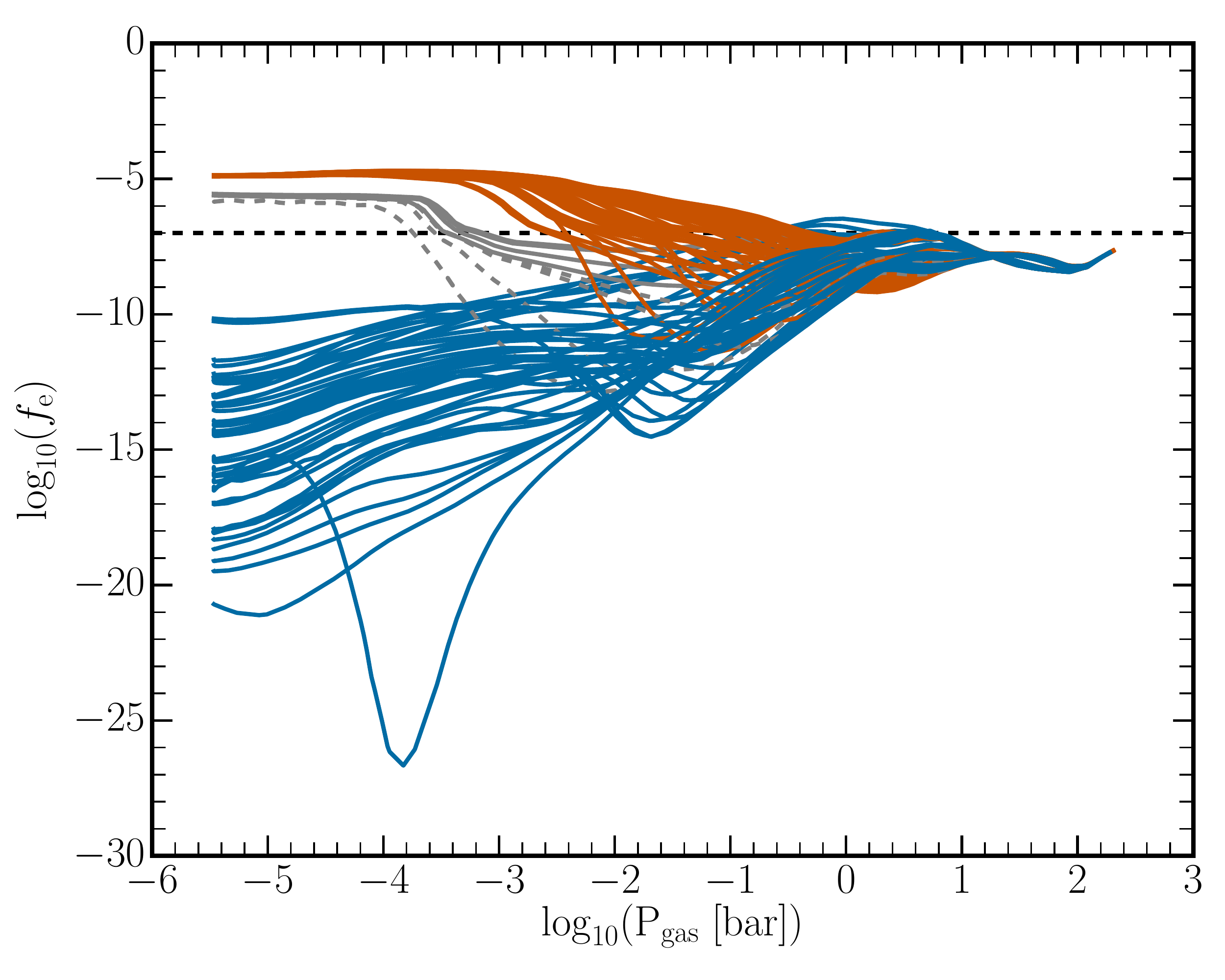}\\
  \includegraphics[width=1.2\linewidth]{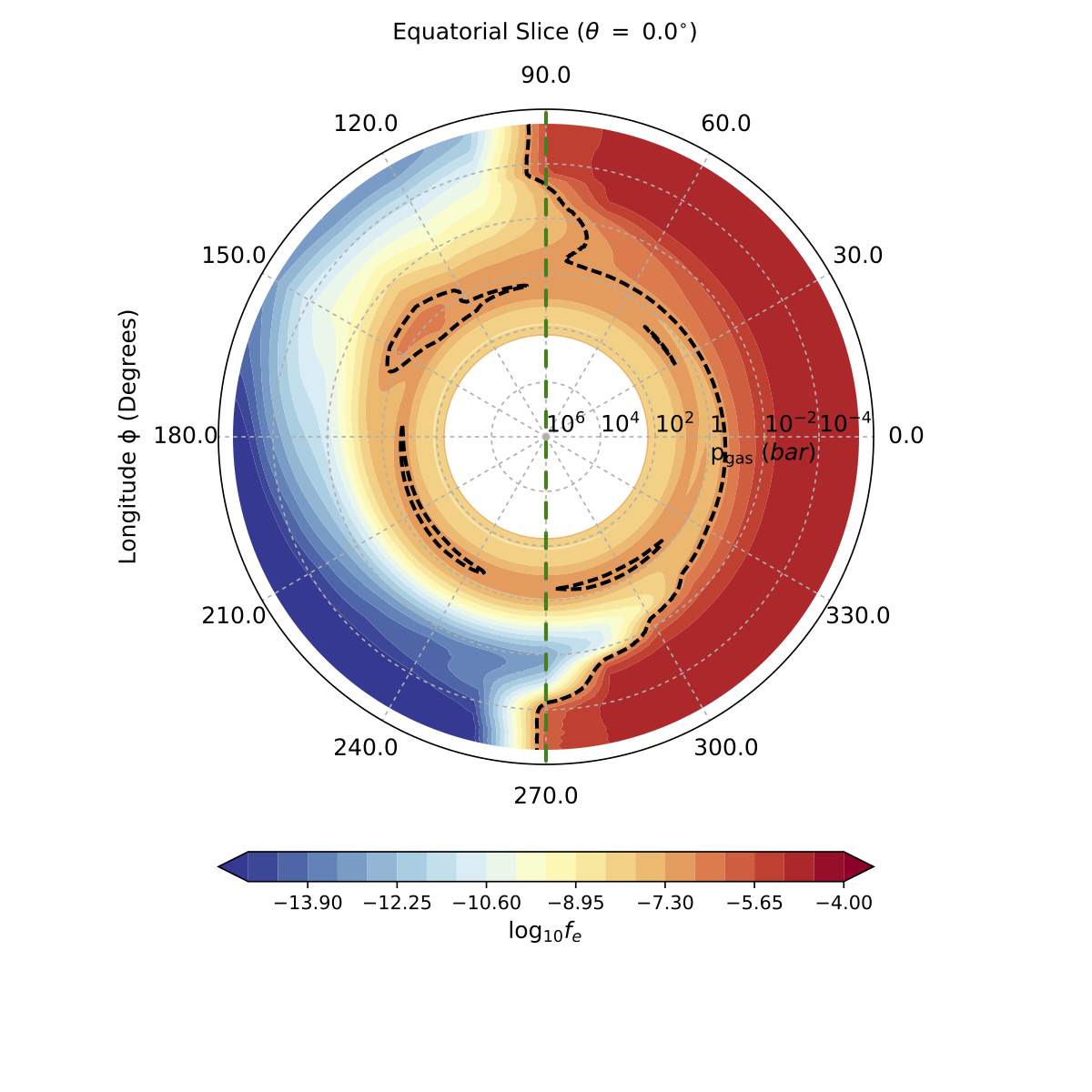}
 &
  \includegraphics[width=1.2\linewidth]{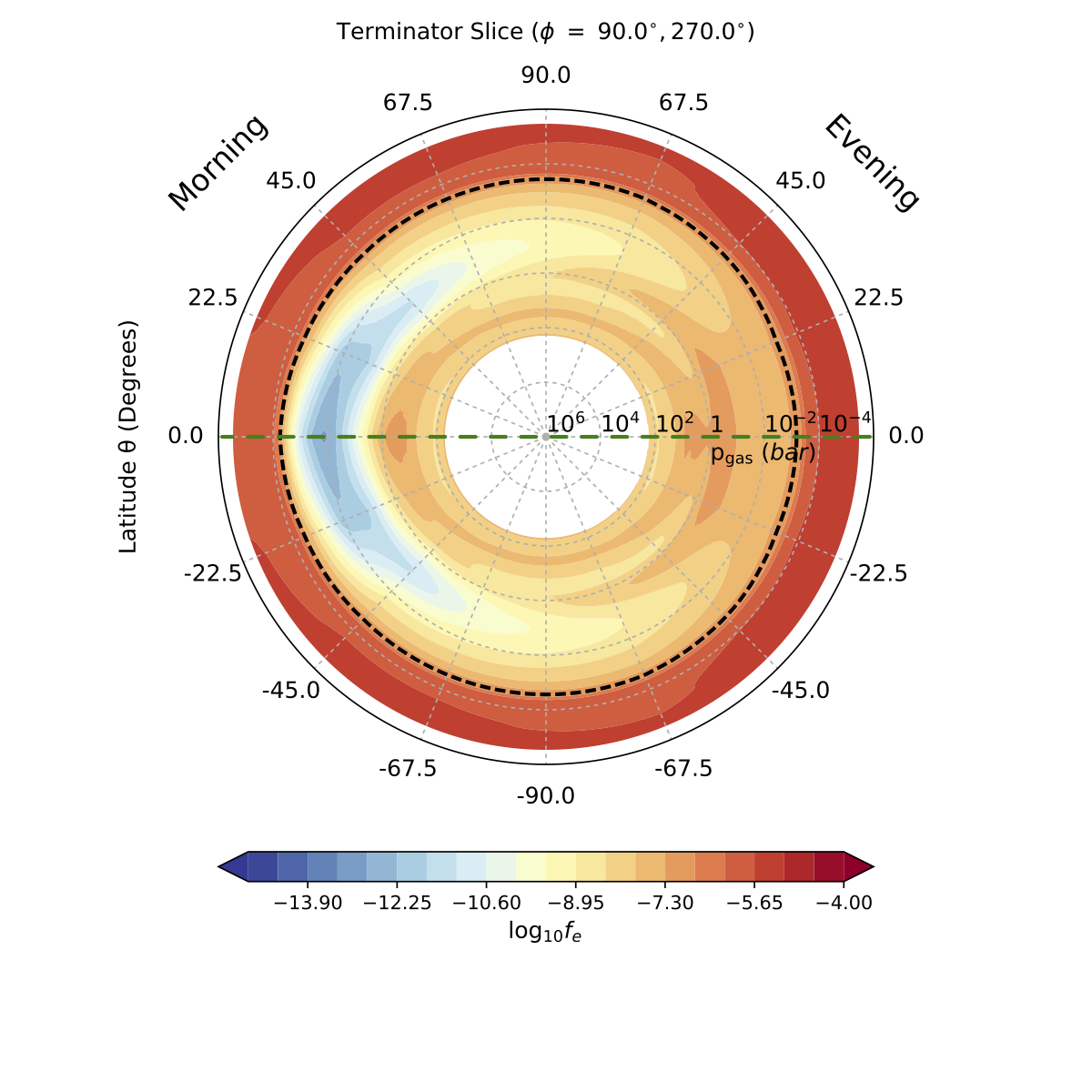}\\*[-1cm]
  \end{tabular}
  \caption{{\bf Top:} HAT-P-7b's ($T_{\rm gas}$, p$_{\rm gas}$) structure (left) and  degree of ionisation, $f_e$, (right) of the 97 1D  profiles. The red lines correspond to the dayside ($\phi=60.0^{\circ} \ldots 0.0^{\circ} \ldots 300.0^{\circ}$). The blue lines are the nightside profiles ($\phi =120.0^{\circ} \ldots 240.0^{\circ}$). The grey solid line corresponds to the evening terminator profile  ($\phi=90.0^{\circ}$). The grey dashed lines are the morning terminator profiles ($\phi=-90.0^{\circ}$or $270.0^{\circ}$). {\bf Bottom:} Equatorial slice plots of the thermal degree of ionisation ($\theta=90^o$ - north, $\theta=-90^o$ - south, left - nightside, right - dayside). The green dashed lines indicate where the two slice plots overlap.  The dashed black line marks the homopause line and illustrates where $f_e=10^{-7}$ in the atmosphere. 
  }
  \label{TEMP}
\end{figure*}

\begin{table}[]
    \centering
    \begin{tabular}{l|l|l}
 \hline
Ion/Atom & Radius  & Reference \\
\hline
\hline
H$_2$ & 136 &  {\tiny \cite{rodriguez2015reference}}\\
Fe$^+$ & 92 &  \cite{ionr}\\
Al$^+$ & 67.5 & \cite{ionr}\\
Si$^+$ & 54 & \cite{ionr}\\
Na$^+$ & 116 &  \cite{ionr}\\
Mg$^+$ & 86  &  \cite{ionr} \\
Ca$^+$ & 114 & \cite{ionr}\\
K$^+$  & 152 & \cite{ionr}\\
Li$^+$ & 90  & \cite{ionr}\\
\hline
    \end{tabular}
    \caption{Ion and H$_2$ radii values used.}
    \label{tab:1}
\end{table}

\subsection{Magnetic Plasma Parameters}\label{iii.iii}

An electrostatic plasma may be magnetised in the presence of a magnetic field.  No direct measurements of magnetic field strength exists for exoplanets, though transit ingress/egress asymmetries may enable to do so in the future (\citealt{2011MNRAS.411L..46V}).  The presence of a planetary magnetic field in combination with the a sufficiently coupled atmospheric gas will affect the global circulation of exoplanet atmospheres (e.g., \citealt{batygin2013magnetically}). This may become of particular interest for ultra-hot Jupiters which exhibit an enhanced thermal ionisation throughout the atmosphere extension on the dayside.

{\it Magnetic coupling:} The comparison of the cyclotron frequency, $\omega_{\rm c}$, and  the collisional frequency, $\nu_{\rm coll}$ results in an expression for a critical magnetic field density, $B_{\rm e}$. $\omega_{\rm c}$ is the angular velocity with which the electrons gyrate around magnetic field lines, and a plasma is magnetised if
$\omega_{\rm ce} \gg \nu_{\rm ne}$, with  $\omega_{\rm ce} =  q_{\rm e} B/m_{\rm e}$ 
For $\omega_{\rm ce} \gg \nu_{\rm ne}$, the critical magnetic field density $B_e$ must be
\begin{equation}\label{befinal}
 B_{\rm e} \gg \dfrac{m_{\rm e}}{e} \sigma_{\rm gas} n_{\rm gas} \left( \dfrac{k_{\rm B} T_{\rm e}}{m_{\rm e}}\right)^{1/2}.
\end{equation}

 When comparing the critical magnetic flux density for electrons, $B_{\rm e}$, to the global magnetic field  strength of a giant gas planet, one can determine to what extent the atmosphere is coupled to the magnetic field. We note, however, that local values can be different. According to the literature, a giant gas planet has a typical magnetic field strength between 3 G and 10 G (\citealt{radii,rodriguez2015reference,2017NatAs...1E.131R}).

\smallskip
{\it Magnetic heating by diffusion:}
The magnetic diffusivity, $\eta$, measures the effect of collisions between the electrons and the neutral particles on the magnetic field. The magnetic diffusivity, $\eta$,  depends on the conductivity, $\sigma$ [S m$^{-1}$] \citep{batygin2013magnetically}.
\begin{equation}\label{eta1}
\eta = \frac{1}{ \mu_0 \sigma},
\end{equation}
\vspace{5pt}
where $\mu_0$ is the permeability of free space. $\eta$ can also be written as a combination of both the decoupled diffusion coefficient and the Ohmic diffusion coefficient, $\eta_{\rm d}$ and $\eta_{\rm ohm}$ respectively, $
\eta = \eta_{\rm d} + \eta_{\rm ohm}$, where
\begin{align}
\eta_{\rm d} = \frac{c^2 \nu_{\rm ne}}{\omega_{\rm pe}^2} && \eta_{\rm ohm} =\frac{c^2 \nu_{\rm ei}}{\omega_{\rm pe}^2}
\end{align}
\vspace{5pt}
 $\nu_{\rm ei}$ is the electron-ion collisional frequency, $\nu_{\rm ei}= \sigma_{\rm ion} n_{\rm ion} \upsilon_{\rm th, e}$ where $\sigma_{\rm ion}$ is the collision cross-section of the ion, $\sigma_{\rm ion} = \pi {r_{\rm ion}}^2$. $n_{\rm ion}$ is the gas-phase ion number density, and $r_{\rm ion}$ is the ion radius 

\begin{figure}
\includegraphics[width=\columnwidth, page=1]{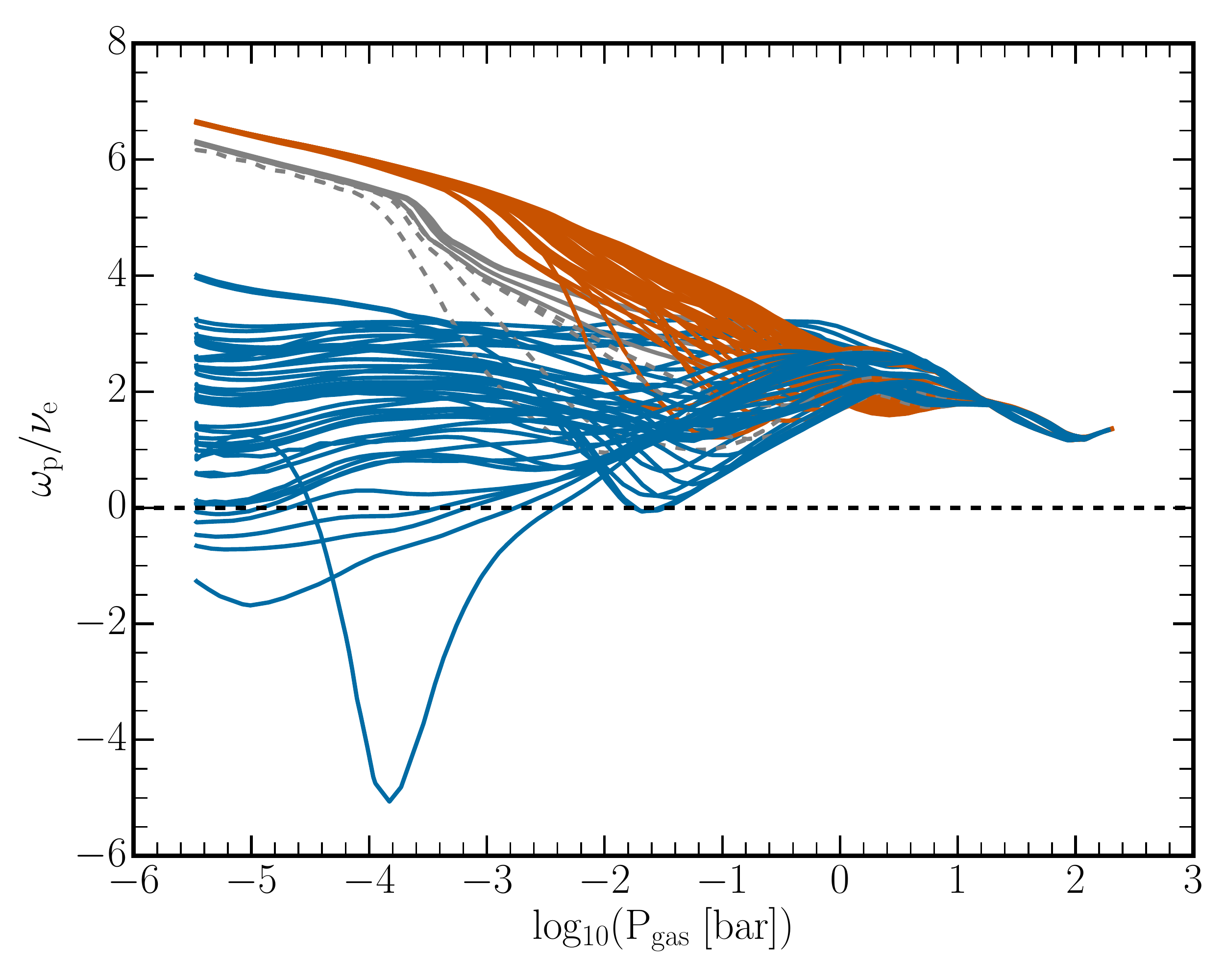}\\ \includegraphics[width=\columnwidth, page=1]{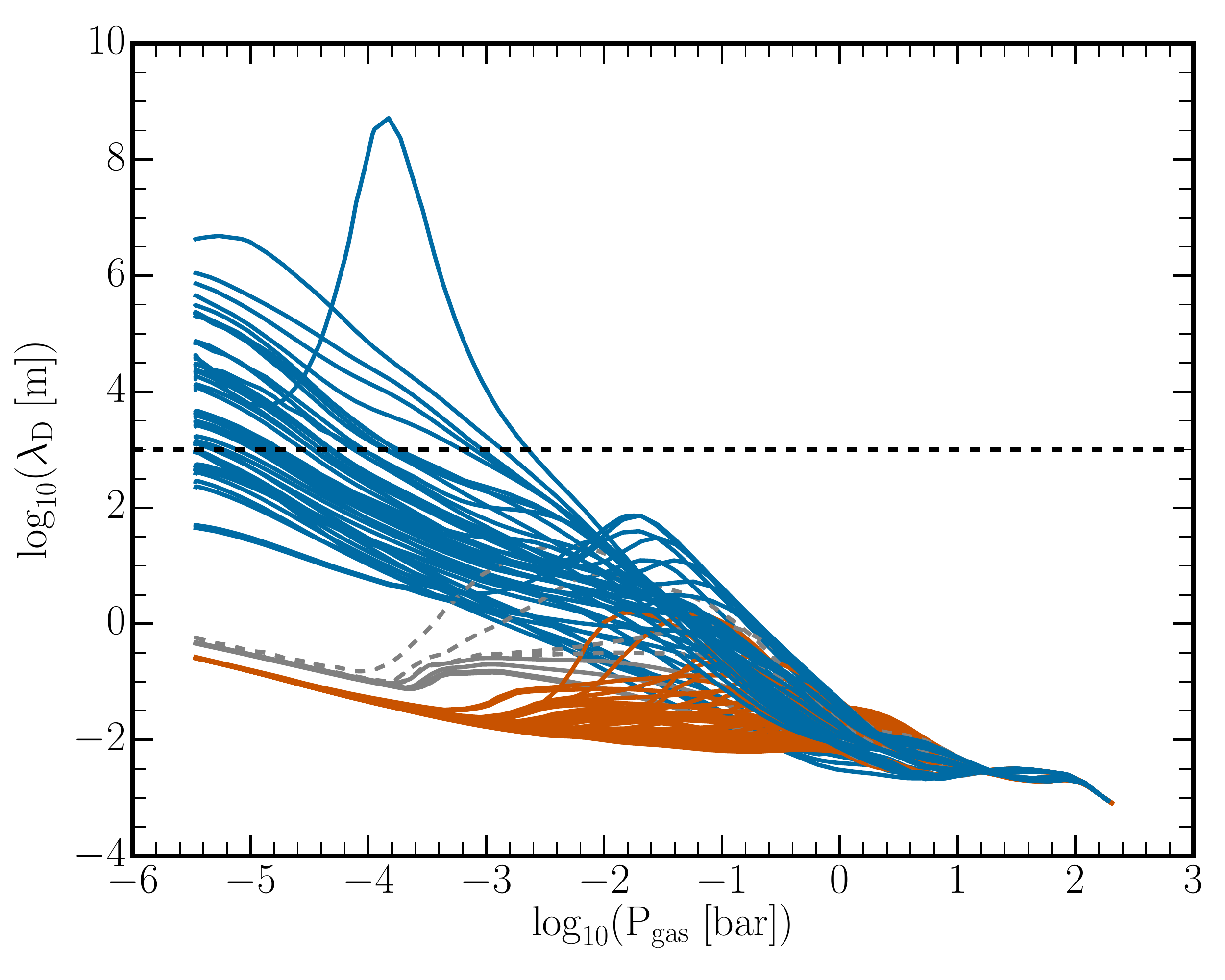}
 \caption{Ratio of plasma frequency of electrons to electron-neutral collisional frequency (top, $\omega_{pe}/\nu_{ne} = 1$ -- black dashed line) and the Debye Length, $\lambda_D$ of all 97 1D profiles. For length scales less than the Debye length, a test charge experiences the influence of a charge imbalance inside a Debye sphere. The horizontal black dashed line at $L=10^{3} $m represents the pressure scale height in the mid-atmosphere for comparison. The red lines correspond to the dayside ($\phi=60.0^{\circ} \ldots 0.0^{\circ} \ldots 300.0^{\circ}$). The blue lines are the nightside profiles ($\phi =120.0^{\circ} \ldots 240.0^{\circ}$). The grey solid lines corresponds to the evening terminator profiles ($\phi=90.0^{\circ}$). The grey dashed lines are the morning terminator profiles ($\phi=-90.0^{\circ}$or $270.0^{\circ}$)}\label{fig8}
\label{omegaDl}
\end{figure}

The full consideration of the conductivity as a tensor is necessary when  considering a plasma.
However, as shown in Fig.~\ref{omegaDl}, the plasma frequency is almost everywhere far greater than the electron-neutral collisional frequency, which means we would not expect a strong plasma behaviour.
Where the previous statement does not hold, for some profiles on the night side, the degree of ionisation is very low (see Fig.~\ref{TEMP}), confirming that we do not have to consider HAT-P-7b’s atmosphere as a fully ionised plasma.
Therefore, as also discussed in \cite{2014ApJ...796...16K} in the case of HD20458b’s atmosphere, the Pedersen and Hall resistivities can be neglected.
 
 Both $\eta_d$ and $\eta_{ohm}$ give insight into the dominance between electron-neutral collisions and electron-ion collisions respectively. It is possible to relate $\eta$ with $\omega_{\rm pe}/\nu_{\rm ne}$: If the latter increases then then magnetic diffusivity  decreases, thus allowing a magnetic field to be generated and transported through fluid motion. This gives opportunity for the magnetic energy to be released into upper layers of the atmosphere in the form of  X-ray or radio emissions \citep{rodriguez2015reference}.

At regions with lower magnetic diffusivity, one expects higher coupling to the magnetic field as there is little loss of field strength due to diffusion. \cite{batygin2013magnetically} suggests that, following magnetohydrodynamical calculations, if the magnetic field of a planet is stronger than the critical value, $B_{\rm e}$, then the upper atmospheric circulation undergoes a transition. The circulation shifts from a state dominated by day-to-night side flows to an azimuthally symmetric one dominated by zonal jets.\\
\vspace{5pt}

\section{Results}\label{results}

\subsection{The Ionosphere of HAT-P-7b}\label{iv.i.i}

Figure~\ref{TEMP} (top left) illustrates the input temperature profiles of the both the day and night side: $\phi=-45.0^{\circ}\ldots0.0^{\circ}\ldots45.0^{\circ}$ and $\phi=-135.0^{\circ}\ldots180.0^{\circ}\ldots135.0^{\circ}$ respectively, as well as the morning terminator ($\phi=-90^{\circ}$) and evening terminator ($\phi=90^{\circ}$). It confirms and highlights the result that the thermodynamical structures are vastly different: there is a $>2500$K change in temperature between the two sides. 
Figure~\ref{TEMP} (top right) demonstrates that thermal degree of ionisation, $f_{\rm e}$, follows a similar trend to the temperature, globally. The dayside profiles all have a similar degree of ionisation of $\backsim 10^{-5}$ at relatively low pressures.  The pressure range where $f_{\rm e}>10^{-5}$ is p$_{\rm gas}<10^{-3}$ bar at the terminators (grey lines  in Fig.~\ref{TEMP}), but ranges up to p$_{\rm gas}<10^{-0.5}$ bar at the dayside.
The nightside fails to reach the threshold of $f_{\rm e} < 10^{-7}$ at such pressures. All of the night profiles are only thermally ionised $> 10^{-7}$ deeper in the atmosphere at $p_{\rm gas} > 10^{-1}$ bar. Both of the terminators are ionised such that $f_{\rm e} > 10^{-7}$ at low pressures but to a lesser extent than the dayside. 

We use the thermal degree of ionisation as a first guidance to locate a possible ionosphere. The ionosphere is defined as the  region of an atmosphere where significant numbers of free thermal electrons and ions are present (\citealt{schunk2009ionospheres}). It is reasonable then, to assume that the ionosphere region would have a degree of ionisation greater $f_e >10^{-7}$.  The ionosphere region will have a lower and upper boundary. The latter will  be at a distance
outside the computational domain of the 3D GCM utilised in this study. We can, however, locate a potential inner boundary where the condition $f_{\rm e} = 10^{-7}$ is met.

 The upper ionosphere  will be determined by external radiation from the host star in the form of XUV, FUV, and SEPs in cosmic rays. \cite{2014ApJ...796...15L} demonstrate that photoionisation of Na and K can reach as deep as 1~bar in HD\,209458b, and that photoionised  Na$^+$ and K$^+$ are also the dominating electron donors for $p_{\rm gas} < 10^{-2}$ bar until H$^+$ kicks in at  $p_{\rm gas} < 10^{-8}$ bar.

Figure~\ref{TEMP} (top right)  indicates that the dayside has a large, deep ionosphere shown by the dashed black line where $f_{\rm e} = 10^{-7}$. The nightside, however, does not. Figure~\ref{TEMP} (bottom) visualises the changing extend of the thermal ionosphere across HAT-P-7b. A  thermal ionosphere is present on the dayside and extends deep in to the atmosphere. Near the terminators, the ionosphere shrinks. On the nightside, 
there is no  thermal ionosphere present for which $f_{\rm e} > 10^{-7}$.

\begin{figure*}
\begin{tabular}{p{8cm}p{8cm}}  
\includegraphics[width=1.05\linewidth,page=1]{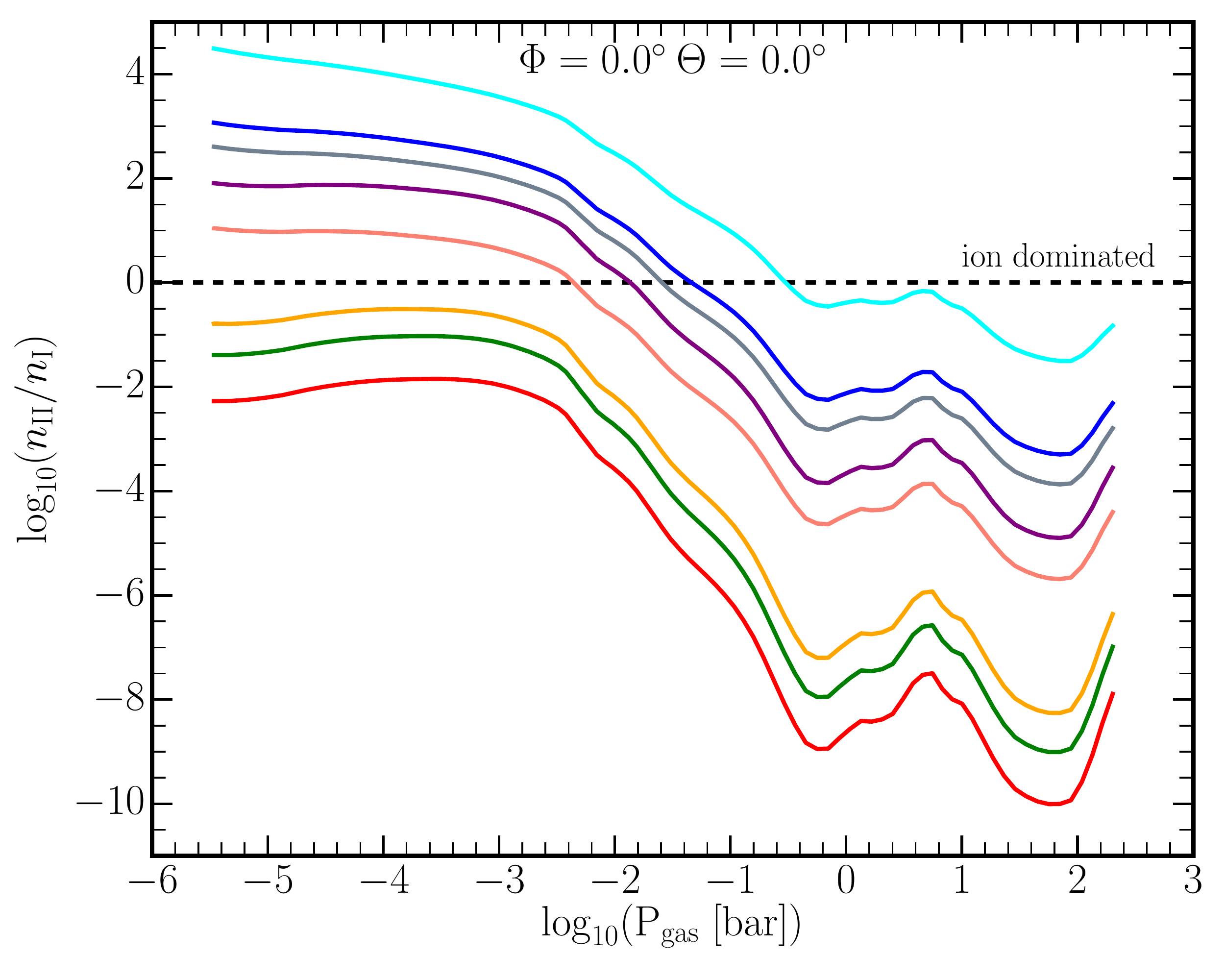} &
\includegraphics[width=1.05\linewidth,page=1]{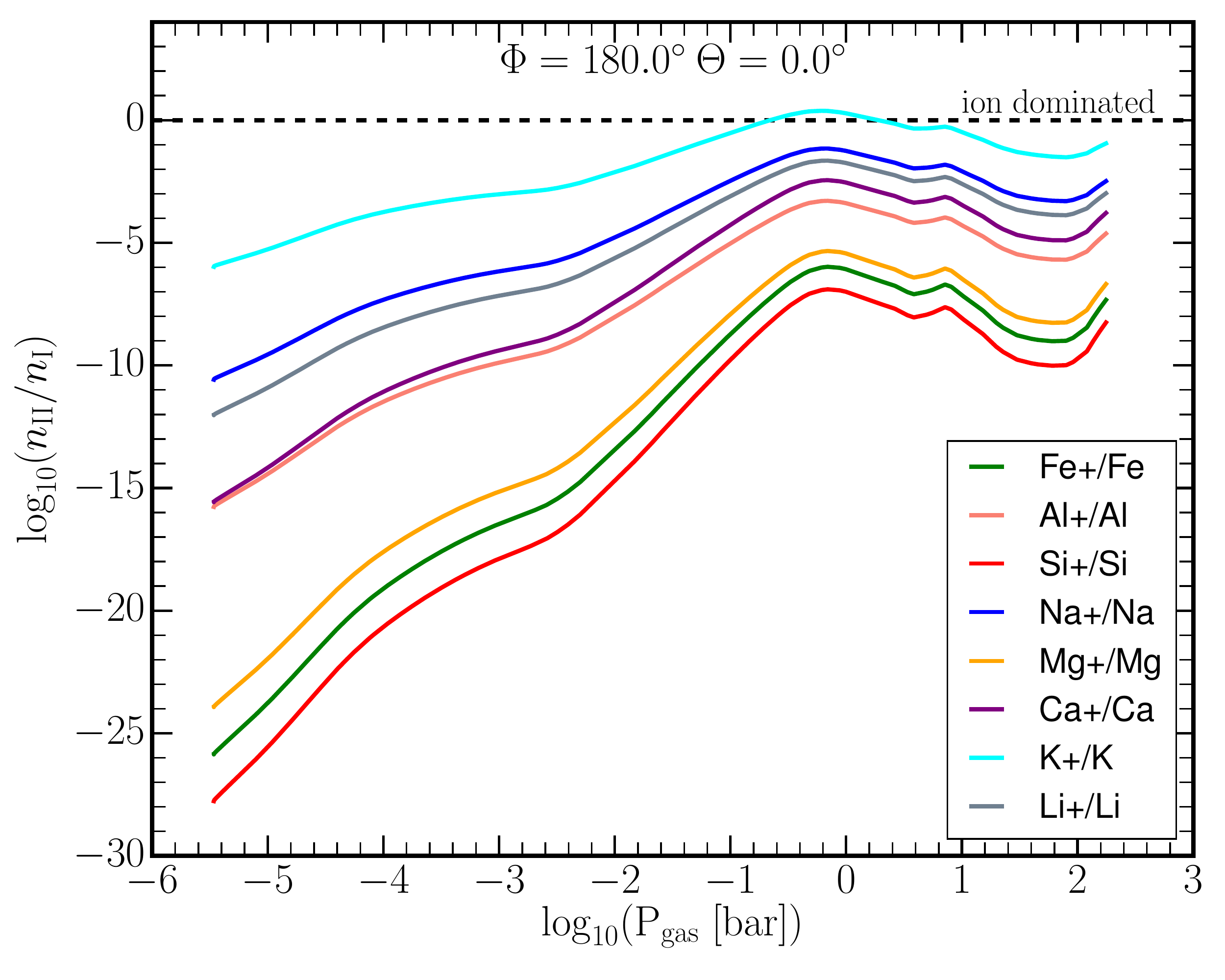}\\
\includegraphics[width=1.05\linewidth,page=1]{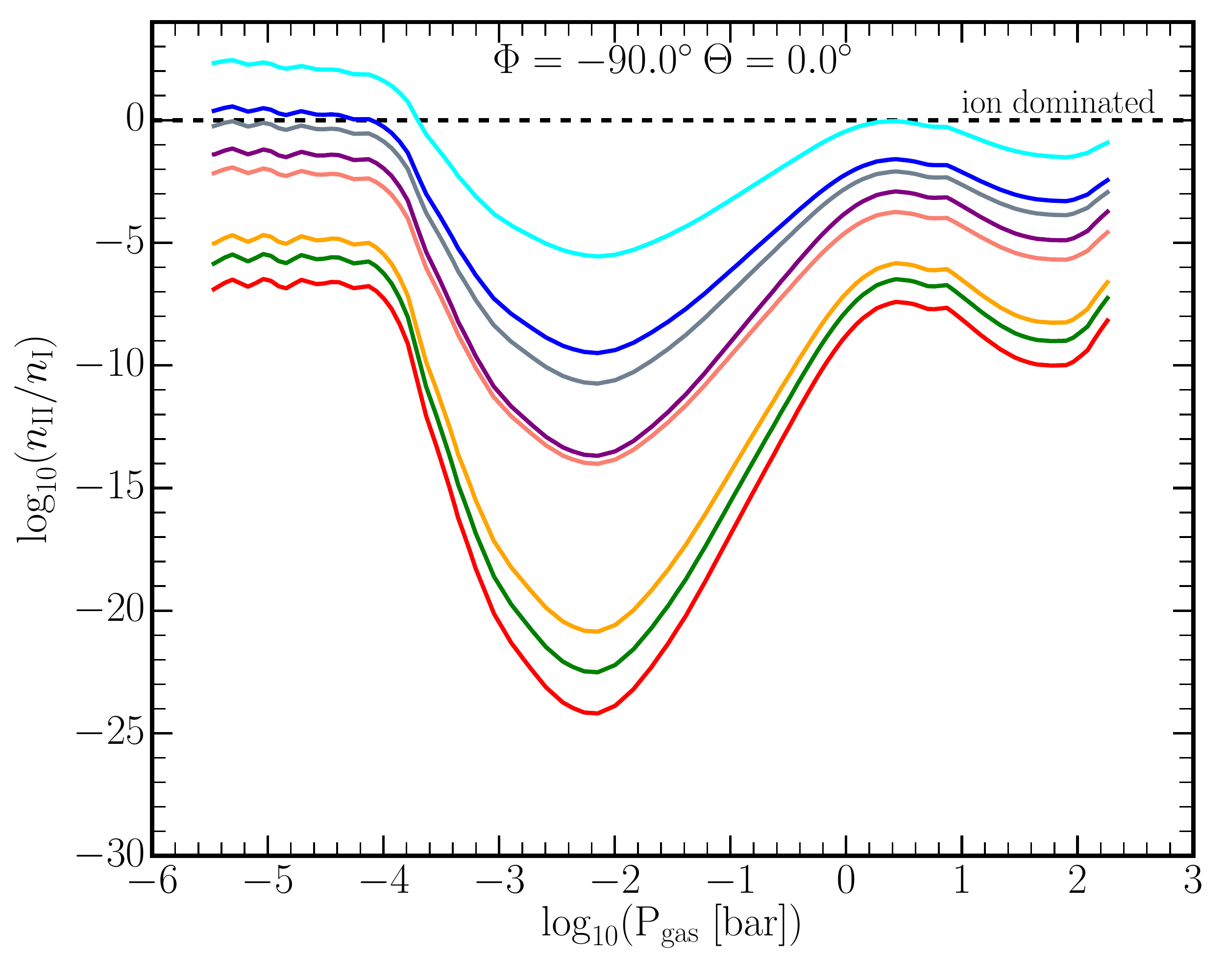}&
\includegraphics[width=1.05\linewidth,page=1]{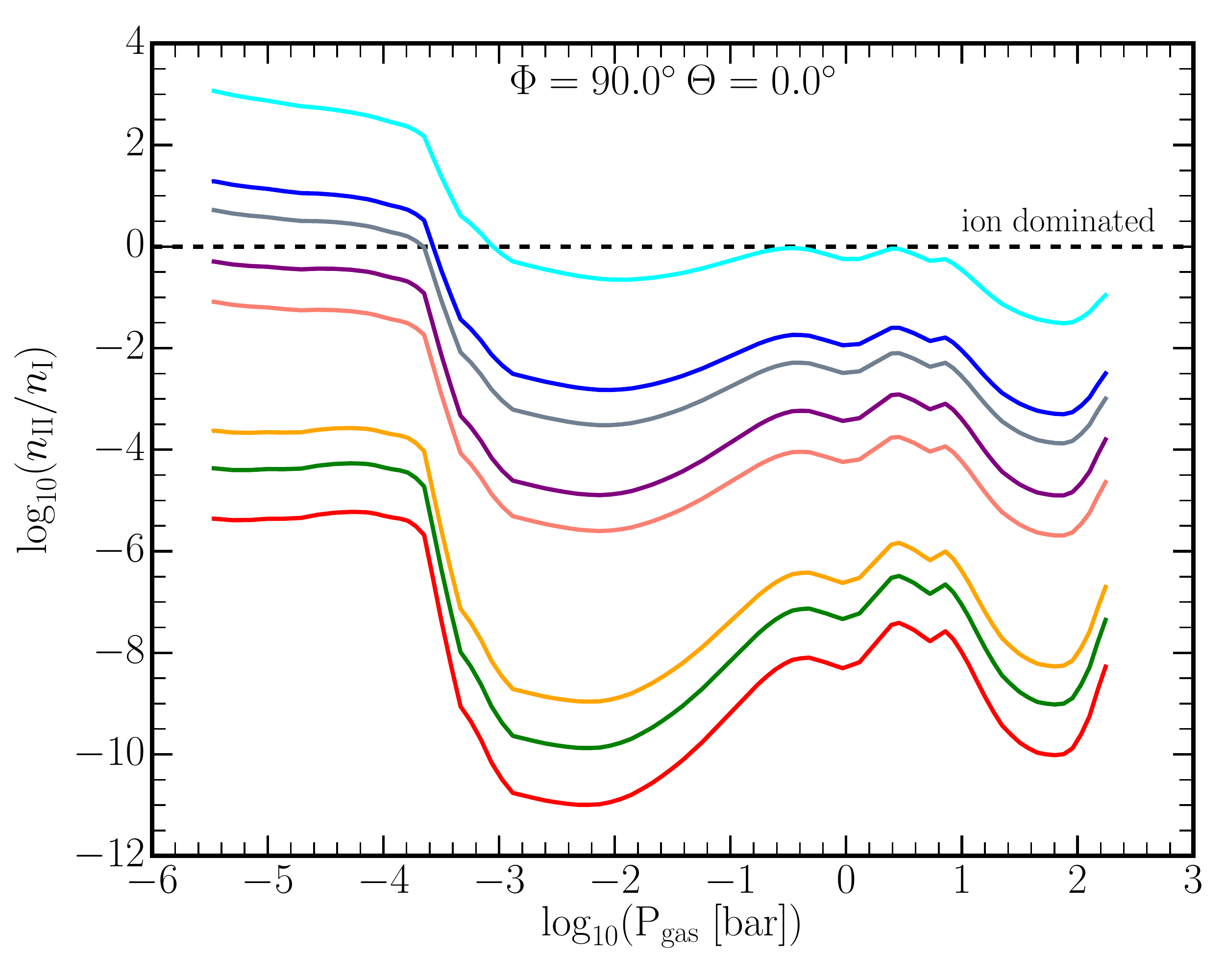}
\end{tabular}
\caption{Dominating ions at the dayside, the nightside and at the two equatorial terminators  of HAT-P-7b ($\phi=90^o$ - evening, $\phi=-90^o$ - morning).  nI is the number density for the gas-phase atom,  nII is the number density for the singly ionised ion in the gas-phase. 
The black dashed lines  indicate where nII/nI$=1$. Above this line,  the element is ionised, and below this line they appear  neutral.}
\label{fig10}
\end{figure*}

Figure~\ref{TEMP} (bottom)  illustrates the location of the lower ionosphere boundary where $p_{\rm gas}=p_{\rm gas}(f_{\rm e} = 10^{-7})$ (black dashed line). 
Figure~\ref{TEMP} (bottom left)  gives an illustration of the thermal ionisation across both the morning ($\phi=-90.0^{\circ}$) and evening ($\phi=90.0^{\circ}$) terminators. We notice that there is a small difference in the location of the inner boundary of the thermal ionosphere between the morning and the evening side.   Figure~\ref{TEMP} (bottom right) suggests that an ionosphere is present in all locations along the terminator of HAT-P-7b. 

The presented results are limited by that  the model atmosphere probes only a certain pressure range and that only thermal ionisation is considered.
An external interstellar radiation field would provide a certain amount of ionisation which could lead to a highly ionised outer shell of the atmosphere that would most likely encircle the whole planet, comparable to what has been demonstrated for brown dwarfs atmospheres by  \cite{rodriguez2018environmental}.  \cite{Barth2020} demonstrate for the giant gas planet HD\,189733b that host star's stellar energetic particles and the stellar XUV amplify the electron production on the dayside. The nightside is affected by thermal and cosmic ray ionisation only. Hence, photoionistaion will amplify the effects discussed in this paper.

\subsection{Possibility and reach of electromagnetic interactions}
\label{iv.i.iii}

In order to investigate where in the atmosphere of HAT-P-7b long-range, electromagnetic interactions  may dominate over short-range binary interactions of an ionised gas we determine where in the atmosphere the plasma frequency exceeds the collisional frequency, i.e. where  $\omega_{\rm pe} \gg \nu_{\rm ne}$. 
Figure~\ref{omegaDl} (top) shows that $\omega_{\rm pe} \gg \nu_{\rm ne}$ for all but a few HAT-P-7b atmosphere 1D profiles probed in our study. Deeper in HAT-P-7b's atmosphere, at gas pressures greater than $10^{-2}$bar, the criteria is met for all profiles tested here. This indicates that deeper in the atmosphere electromagnetic interactions dominate, globally. This is in accordance with expectation according to the higher number of thermally ionised electron with higher temperatures, hence deep inside the atmosphere and on the dayside. It is, nevertheless, remarkable that long-range, electromagnetic interactions should also be expected to affect the atmosphere on most of the nighside of HAT-P-7b. Our finding implies that MHD processes may not be confined to small scales only.

 We note that \cite{2014ApJ...796...16K} find that $\omega_{\rm pe} \gg \nu_{\rm ne}$ for p$_{\rm gas}<10^{-2}$bar for the dayside of HD\,209458b based on the 1D $(T_{\rm gas}, p_{\rm gas})$-profiles from \cite{2011ApJ...737...15M}. Hence, the inner atmospheric part of HD\,209458b the electrons and ions equilibrate with the neutral atmosphere, but not so in the inner atmosphere of HAT-P-7b according to the $(T_{\rm gas}, p_{\rm gas})$-profiles applied here.

\smallskip
To investigate the reach of electrostatic interactions, the Debye Lengths, $\lambda_{\rm D}$, across HAT-P-7b is explored. We compare these Debye lengths to an atmospheric pressure scale length of $L=10^{3} $m  (e.g., \citealt{helling2011ionizationTWO}).  Figure~\ref{omegaDl} (bottom) shows  how $\lambda_{\rm D}$ changes depending on the local atmospheric gas pressure and how it changes across the globe of HAT-P-7b. The Debye length (Eq.~\ref{eq4}) is related to the plasma frequency, $\omega_{\rm pe}$, as
\smallskip
\begin{equation}\label{LAMBDA}
{\lambda_{\rm D}}^2 = \dfrac{k_{\rm B} T_{\rm e}}{{\omega_{\rm pe}}^2 m_e}
\quad\quad\quad \Rightarrow {\lambda_{\rm D}}^2 \sim \frac{T_{\rm e} }{{\omega_{\rm pe}}^2},
\end{equation}
hence, a maximum in $\omega_{\rm pe}$ must coincide with a minimum in $\lambda_{\rm D}$ as shown in Fig.~\ref{omegaDl}.

\smallskip
If $\lambda_{\rm D} \ll L$ a plasma is quasi-neutral and the ionised gas region exhibits plasma behaviour. Figure~\ref{fig8} highlights that $\lambda_{\rm D} \ll L$ occurs throughout the whole atmosphere of HAT-P7b on the dayside and both terminators. The nightside, however, has many profiles in which $\lambda_{\rm D} > L$ at  low pressures. This, in conjunction with the results in Section~\ref{iv.i.i}, suggests that the upper atmosphere on the nightside of HAT-P-7b may not exhibit plasma behaviour in the computational domain probed by our study. In this region the thermal electron density is low, causing an increase in Debye length. Lower in the atmosphere on the nightside, however the electron density is much greater and so the Debye length is smaller and the criterion is met. The Debye length investigation thus confirms the earlier conclusions.

\subsection{Dominating Ions}\label{iv.ii}
For completeness, we present  the most abundant local gas ions to ascertain which are the dominating electron donors on HAT-P-7b, and hence 
link the plasma parameters  explored in Sects.~\ref{iv.i.i} and ~\ref{iv.i.iii} to the local chemistry in the atmosphere of HAT-P7b. In Sect.~\ref{detectability}, we show which of these ions can be used a spectroscopic tracers for the ionosphere of an Ultra-Hot Jupter like HAT-P-7b.
Figure~\ref{fig10} shows the ratio of atomic number density, nI, and ionic number density, nII,  as a function of local gas pressure for the four important points, stellar ($\phi, \theta$)=(0$^\circ$,0$^\circ$)  and anti-stellar point ($\phi, \theta$)=($180^\circ$,0$^\circ$), morning ($\phi, \theta$)=(90$^\circ$,0$^\circ$) and evening ($\phi, \theta$)=(-90$^\circ$,0$^\circ$)  terminator. These plots probe the dominating ions on the dayside, nightside and two terminators.

Figure~\ref{fig10} demonstrates that the thermal ionisation of the dominating ions K$^+$, Na$^+$, Li$^+$, Ca$^+$, Al$^+$ is considerable at lower pressures at the substellar  point. Here, ions are  $10^2\,\ldots\,10^4\times$ more abundant than the neutral atoms of these species. Fe$^+$,  Al$^+$, Si$^+$, Na$^+$, Mg$^+$, K$^+$, Ca$^+$ and Li$^+$ are the dominating thermal electron donors on the dayside. Species that have sufficiently low ionisation energies and sufficiently high number densities will contribute most effectively to the thermal degree of ionisation. Hence  overall, K$^+$, Na$^+$, Li$^+$, Ca$^+$, Al$^+$ provide the majority of electrons in the atmosphere of HAT-P-7b globally.
Ca$^+$  and also Fe$^+$ have been observed in Ultra-hot Jupiters \citep{casasayas2019atmospheric}, but no such observations have yet been conducted for HAT-P-7b.

The small effect of thermal ionisation  on the nightside appears in Fig.~\ref{fig10} (top right) as that the atoms dominate over the ions in number of a species. The terminator regions do show a slight asymmetry with respect to their degree of ionisation (see Fig.~\ref{TEMP}), which is, however, more pronounced when looking at the ion abundances. Figure~\ref{fig10} (bottom) suggest also a chemical asymmetry between the terminator regions: The evening terminator ($\phi, \theta$)=($90^\circ$,0$^\circ$) would suggest $n(Li^+)>n(Li)$ but $n(Li^+)\approx n(Li)$ at the  morning terminator ($\phi, \theta$)=($-90^\circ$,0$^\circ$). A substantial increase of K$^+$ is suggested for the morning terminator. We note, however, that these need to be revisited based on the next generation of 3D GCM models as the handling of the terminator radiative transfer (perpendicular irradiation) may affect the temperature structure.

We note that any process that impacts the element abundance of the dominating electron donors - cloud formation for example - will affect the electric state of HAT-P-7b's atmosphere, and hence, also the potential coupling to a magnetic field.
This is illustrated in the differences between the morning and evening terminators, which have similar degrees of ionisation (Figure \ref{TEMP}, right). But the drop between $10^{-4}$ and $10^{-1}$ bar seen for the morning terminator in Figure \ref{fig10} (bottom left) is substantially larger than for the evening terminator (note different axis range between sub-figures). This is associated with the region in which clouds form on the morning terminator, something which does not occur on the evening terminator \citep{hatp7b1}.

\begin{figure}
  \includegraphics[width=90mm,page=1]{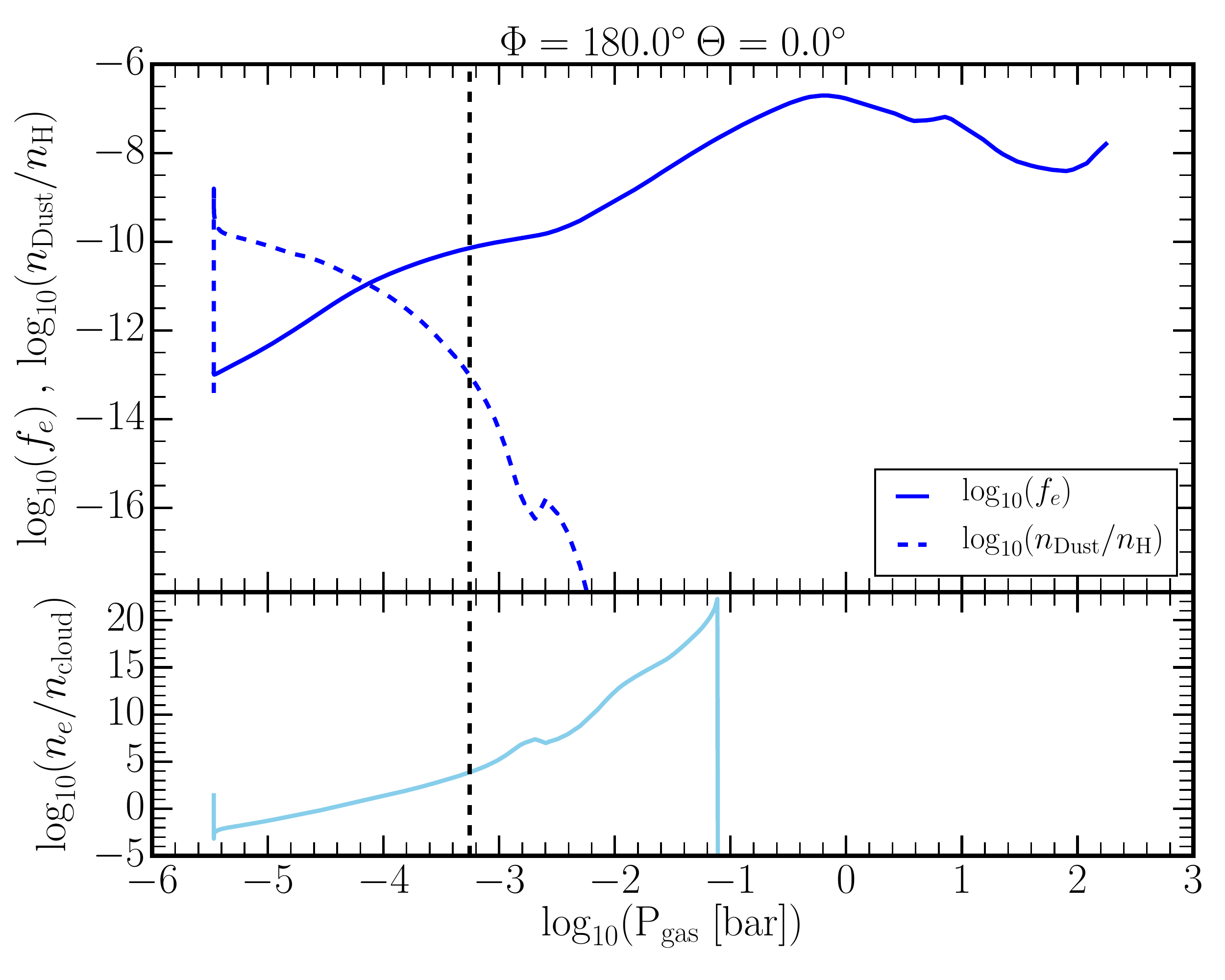}
  \includegraphics[width=90mm,page=1]{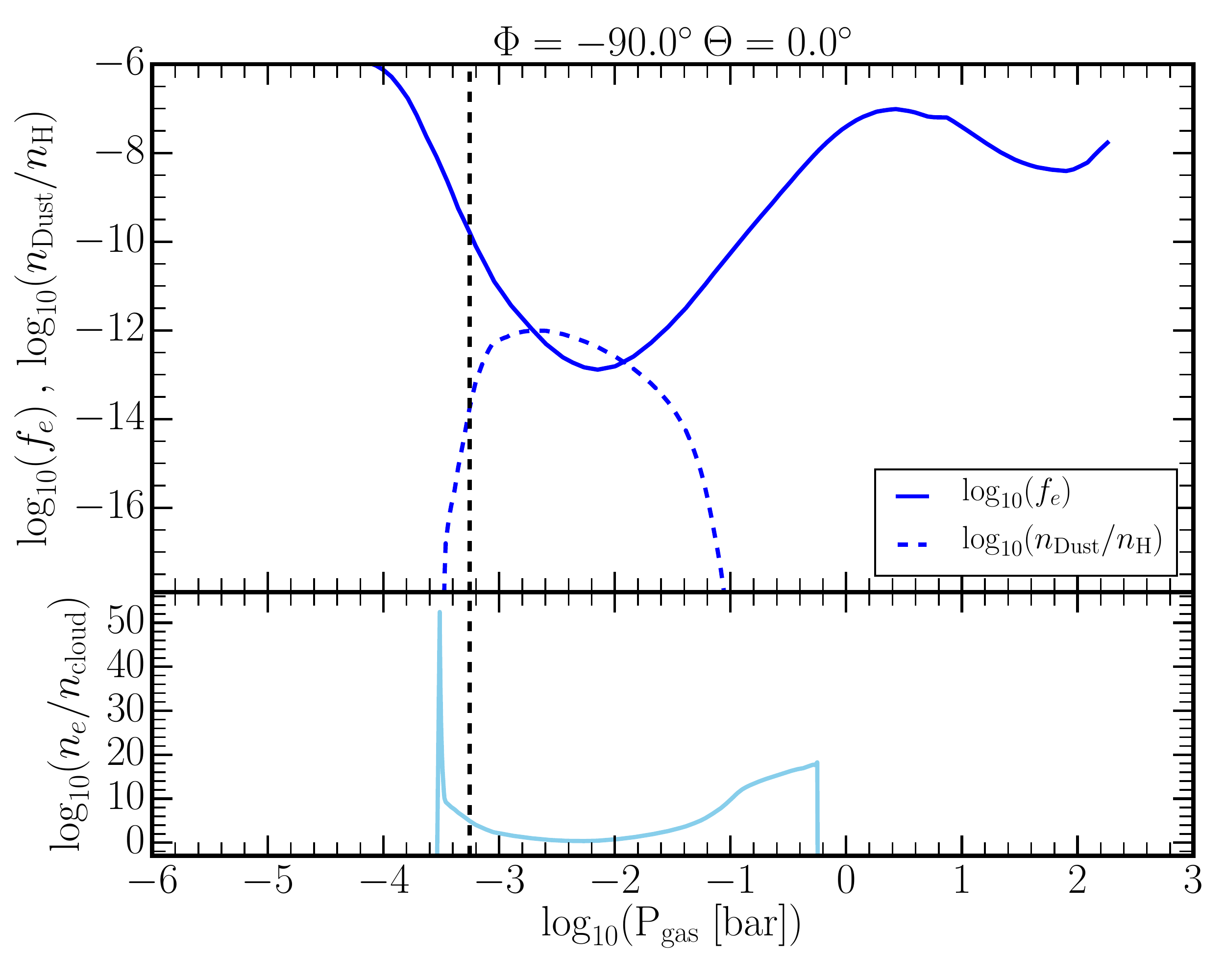}
\caption{Possible discharges in the atmosphere of HAT-P7b on the nighside (top figure) and at the morning terminator (bottom figure). Top panels: Degree of ionisation, $f_{\rm e}$, and the dust number density per hydrogen, $n_{\rm d}/n_{\rm H}$. Bottom panels: Ratio of electron number density, $n_{\rm e}$, to the number of cloud particles, $n_{\rm d}$. 
For $n_{\rm e}/n_{\rm d} >1$, the atmosphere is  in the electron dominated regime.
The dotted vertical line represents $p_{\rm gas}=5.56\times10^{-4}$bar, the pressure limit below which electric field breakdown may occur. To the left of this line, the clouds are susceptible to large-scale discharges.} 
\label{fig11}
\end{figure}


\subsection{The possibility of discharge events on the morning terminator and the nightside of HAT-P7b}\label{iv.iii}

The investigation of the  individual plasma parameters led us to suggest the presence of globally asymmetric ionosphere that reaches deep into the atmosphere of HAT-P7b on the dayside. In the following, we explore implications for the emergence of discharge processes like lightning in the cloudy parts of HAT-P-7b.

Clouds in exoplanet atmospheres can be susceptible to discharge event (\citealt{2016SGeo...37..705H,2016MNRAS.461.3927H,2016MNRAS.461.1222H,2019JPhCS1322a2028H}), similar to what is known for Earth, Jupiter and Saturn. Local discharge events, lightning or sprites for example, have the ability to affect the local gas chemistry and to further ionise the gas (e.g. \citealt{2014ApJ...784...43B}). 

The emergence of lightning in a cold gas is triggered by streamer events which are unstable ionisation fronts that self-amplify.  Their emergence requires seed electron in the gas phase and a sufficient electrostatic potential difference within clouds to accelerate these electron such that they ionise enough of the ambient gas in order to set off a self-sustained ionisation front. Hence, if  an ensemble of  cloud particles is sufficiently charged such that an electrostatic field breakdown can occur and if enough free electrons are available in the gas phase, then a streamer may trigger a lightning event.
Therefore, comparing the degree of thermal ionisation, $f_e$ (Eq.~\ref{eq1}), with the number of cloud particles, $n_{\rm d}$,  enables a first indication for where lightning may be initiated through the formation of streamers  (\citealt{2011ApJ...727....4H,2016PPCF...58g4003H}). If a substantially larger number of electrons than cloud particles are present, it can safely be assumed that  seed electrons remain in the gas phase.

The degree of thermal ionisation, as an indicator for where charges are available in the atmosphere, is compared with the number of cloud particles per hydrogen, $n_{\rm d}/n_{\rm H}$ in Fig.~\ref{fig11} (top panels). Some of the free electrons (or charges) will attach to the surface of the cloud particles and contribute to  charging the cloud particles electrostatically;  other mechanisms like triboelectric charging through relative particle motions may also contribute to the cloud particle charging. Martian dust grains, for example, may acquire charges of $10^3\,\ldots\,10^5e$ collisionally 
(\citealt{2012P&SS...60..328M}). This suggest that addition gas-phase charge pick-up is unlikely in atmospheric regions of high cloud particle number density and strong hydrodynamical winds. \cite{2012P&SS...60..328M} further point out that the particle charge load will also be affected by the ambient gas pressure because of the built-up of a cloud particle surface potential. Frictional charging is thought to be more efficient in dense atmospheric environments.  
Those atmospheric regions where  $n_{\rm e}/n_{\rm d} >1$ (Fig.~\ref{fig11}, bottom panels)  and where also a substantial number of cloud particles are present, will be suitable for large-scale lightning discharge to be triggered if an electrostatic field breakdown is possible.  In case of small cloud particle numbers, local coronal discharges may still occur but it may not develop into large-scale lightning. 

We can estimate the local gas pressure below which an electrostatic field breakdown may occur, but we do not address the emergence of the electrostatic field in this paper. The minimum potential difference (Stoletow point) required for a electrostatic field breakdown in a gas is determined by  $p_{\rm gas} d = 2.509$ Torr cm for a H$_2$-dominated atmosphere and a breakdown distance $d$ (see \cite{helling2013ionizationTHREE,2017GeoRL..44.2604K, 2019Icar..333..294K} for more details).
For breakdown distances of $d>6\,$cm the streamer mechanism dominates (over small-scale coronal discharges) and, hence, may trigger a large-scale discharge in form of lightning  (\citealt{raizer1991gas}). We utilize this value to derive the local gas pressure $p_{\rm gas} = 5.56\times10^{-4} $ bar below which an electric field breakdown would be possible if seed electrons are available (Fig.~\ref{fig11}, vertical dashed line).

Figure~\ref{fig11} suggests that the morning terminator (($\phi, \theta$)=($-90^\circ$,0$^\circ$)) has a thin cloudy region where corona-like discharges may occur if the cloud particles are sufficiently charged, just to the left of the vertical line of $p_{\rm gas} = 5.56\times10^{-4} $ bar.   But no lightning-precursor can be expected here because of too few cloud particles being present.
The situation may be more favourable on the nightside of HAT-P-7b where a larger number of cloud particles is present in the low-pressure atmosphere where $p_{\rm gas} < 5.56\times10^{-4} $ bar, and still $n_{\rm e}/n_{\rm d} >1$. We note that the pressure threshold for streamers to trigger lightning will move into the higher atmosphere with lower pressures if a larger discharge length appears more realistic.

\cite{2014ApJ...784...43B} report an increased amount of CH and C$_2$H at the expense of CH$_4$ in the gas affected by the increased temperature in the lightning channel environment. Simultaneously, a drop in H$_2$O and other metal oxide molecule  would appear.  A potential photometric flickering has so far proven difficult to detect on brown dwarfs (\citealt{2020MNRAS.495.3881H}).

\subsection{The magnetised atmosphere of HAT-P7b}\label{iv.iv}
\begin{figure}[H]
\includegraphics[width=\columnwidth,page=1]{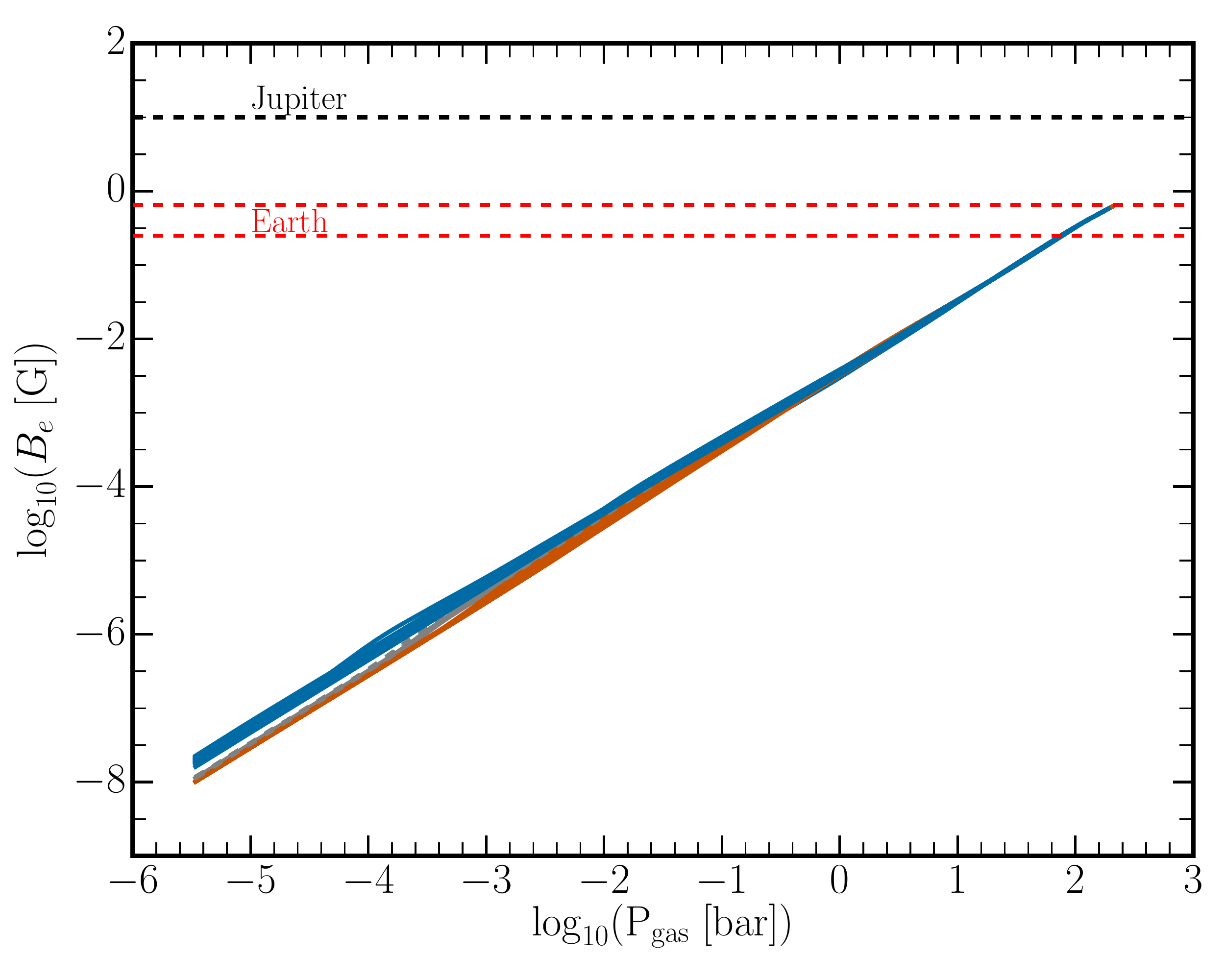}
\caption{\footnotesize Critical magnetic flux density required for electrons, $B_e$, to be magnetically coupled to an external magnetic field on HAT-P-7b. The black dashed line indicates B=10 G which is a typical magnetic field strength for giant gas planets, the red dashed lines are values for the Earth magnetic field strength. If $B > B_e$ then $\omega_{ce} \gg \nu_{ne}$ is fulfilled and the atmosphere is magnetised. All line colours are as in Fig.~\ref{fig8}.
}\label{fig13}
\end{figure}

We explore implications for the magnetic coupling of the atmosphere of HAT-P-7b to a global magnetic field.
 We estimate a critical magnetic flux density (Eq.~\ref{befinal}) for the atmosphere of HAT-P-7b that is required to enable a magnetic coupling of the partially ionised gas to a global magnetic field within our computational domain. Figure~\ref{fig13} suggest that for a global magnetic field of 10\,G (which is somewhat larger that the 6\,G suggested in \cite{2017NatAs...1E.131R}), the whole atmosphere could be affected by the magnetic coupling. Only if a global magnetic field drops below $10^{-3}$\,G, half of the atmosphere would be less or unaffected by magnetic coupling.  This is considerably less than the typically inferred field strength of hot Jupiters (\citealt{christensen2009energy,batygin2013magnetically}). Therefore, moderate global magnetic field strength seem sufficient in order for a magnetosphere to establish on HAT-P-7b which then could interact with the host stars magnetised wind. It remains to be seen if such an interaction would shield (confine) or remove (not confine) the planetary atmosphere (eg. \citealt{2020MNRAS.494.2417V}).

\begin{figure*}
\centering
  \includegraphics[width=90mm,page=1]{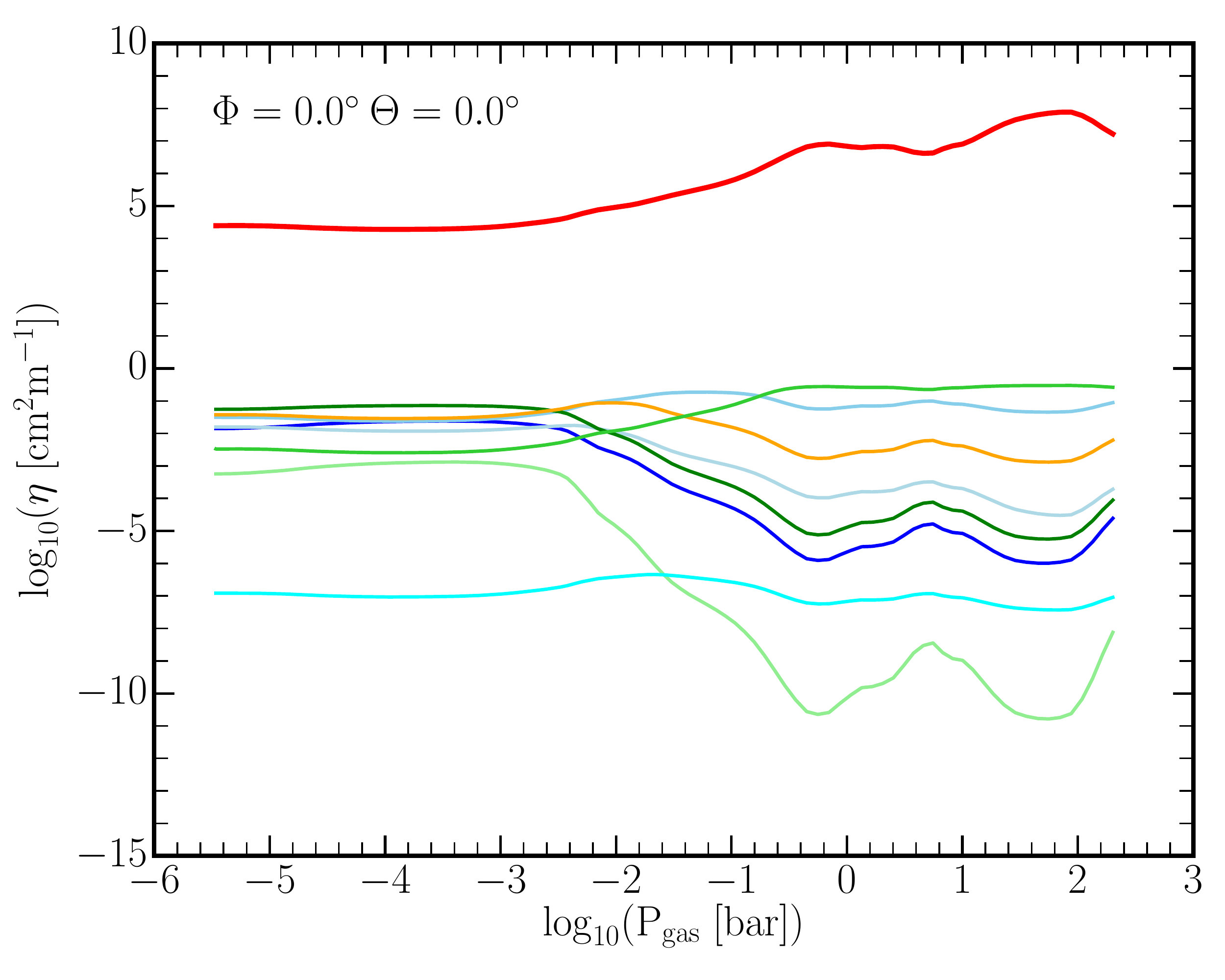}
  \includegraphics[width=90mm,page=1]{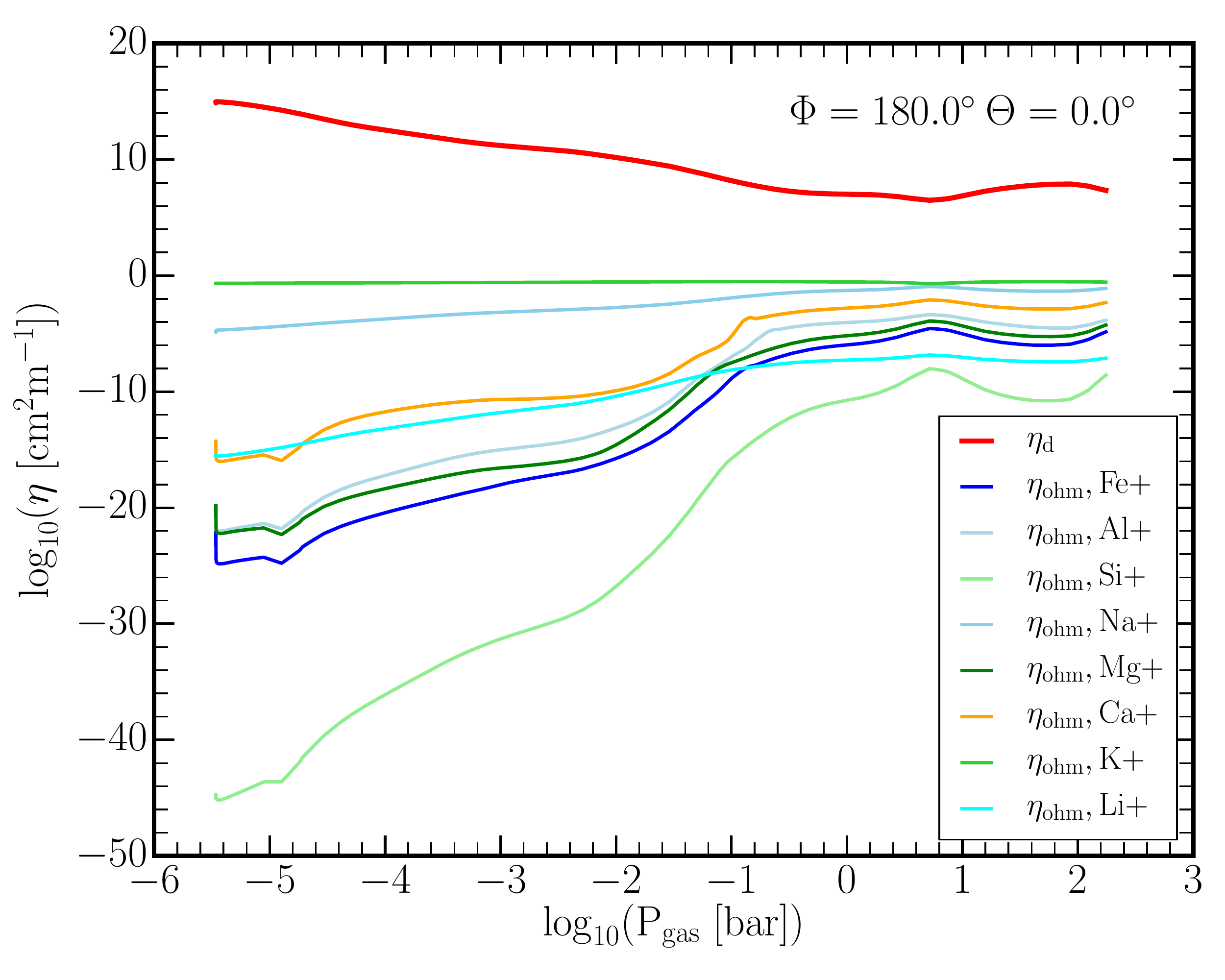}\\
  \includegraphics[width=90mm,page=1]{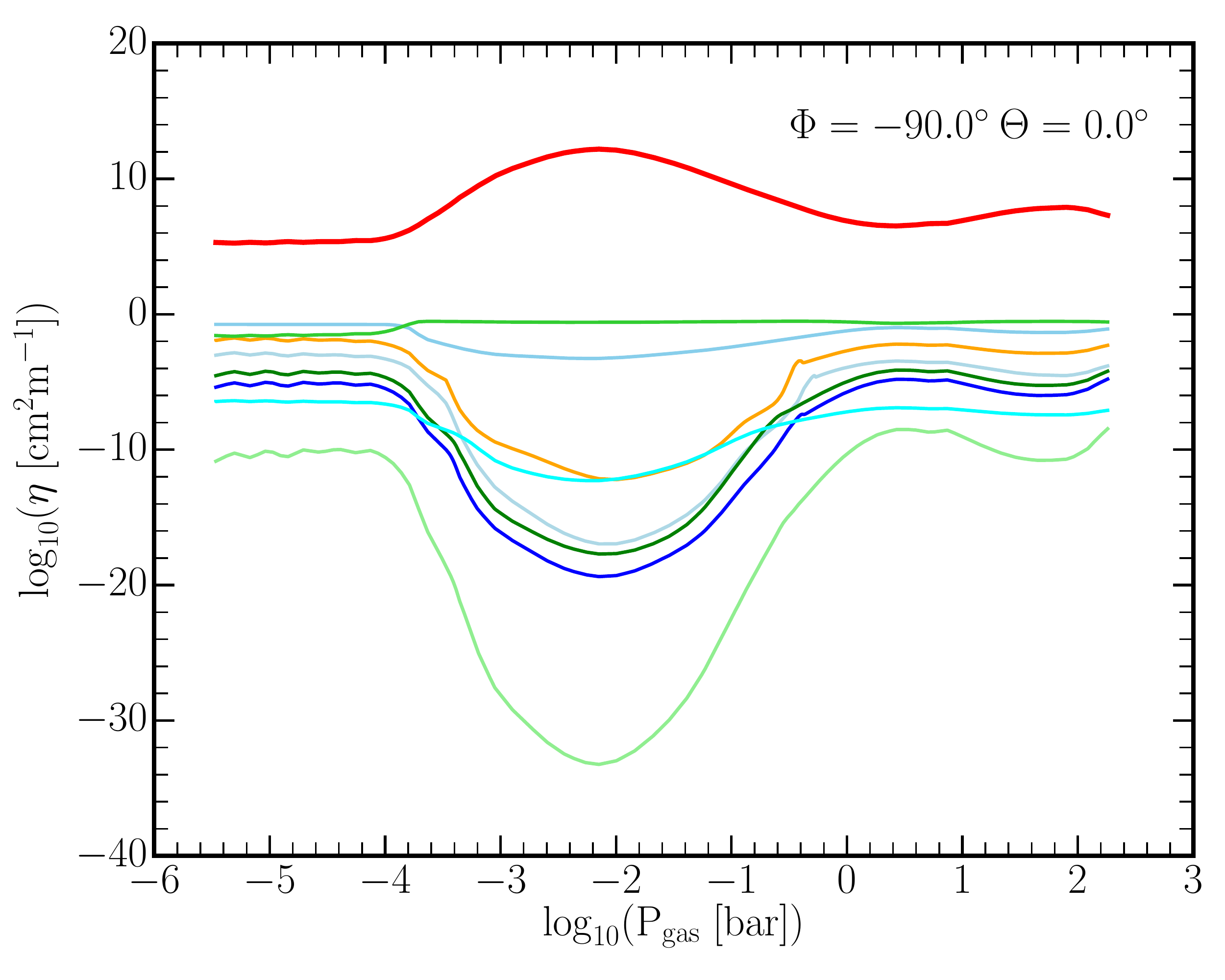}
  \includegraphics[width=90mm,page=1]{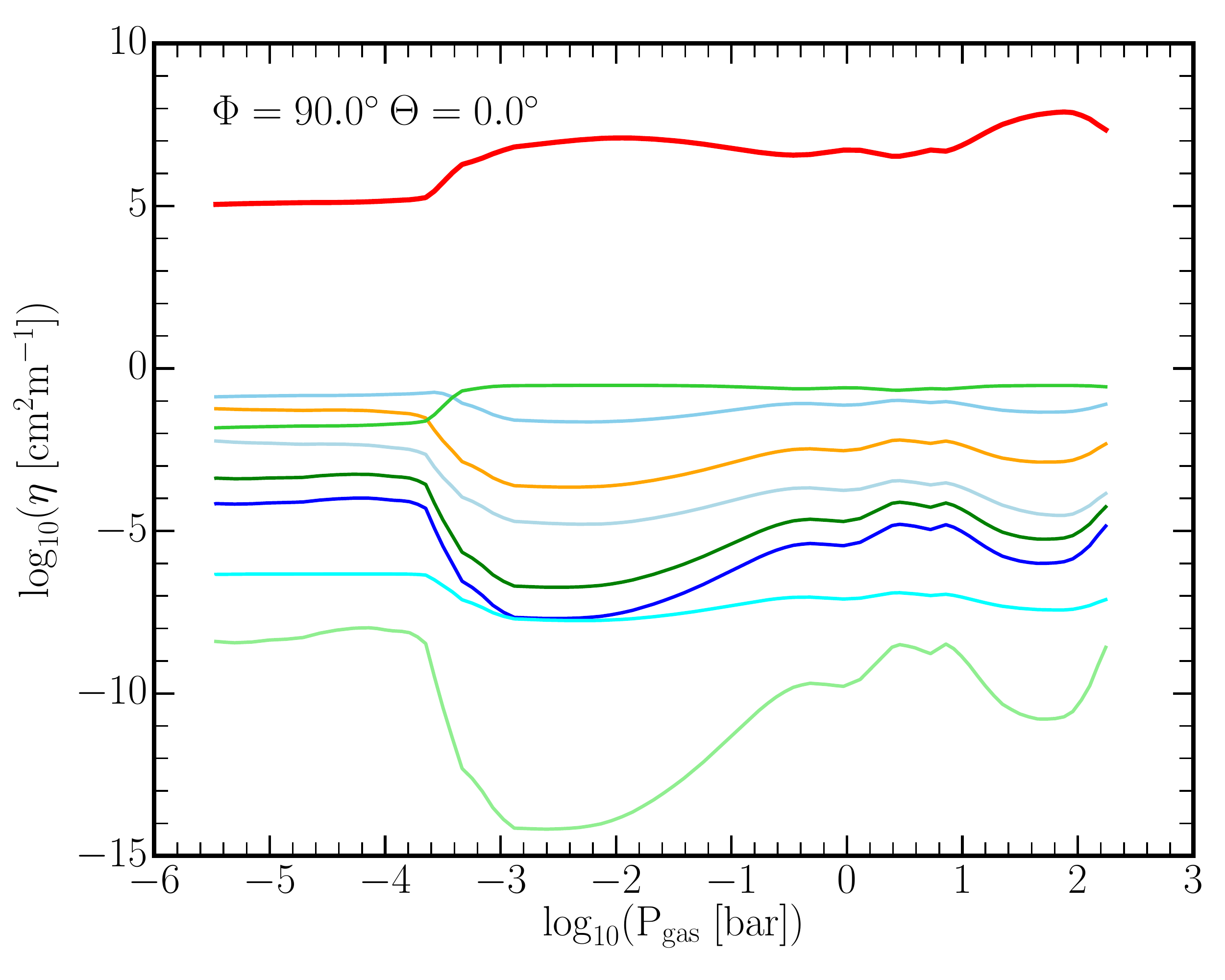}
\caption{\footnotesize The decoupled diffusion coefficient, $\eta_d$, and the Ohmic diffusion coefficient, $\eta_{ohm}$, for the dominating thermal electron donors. The Ohmic diffusion coefficients are all much smaller than the decoupled diffusion coefficient. The electron-ion interactions ($\eta_{ohm}$) are not significant when compared to the electron-neutral interactions ($\eta_d$) upon consideration of thermal ionisation.}\label{fig12}
\end{figure*}

Figure \ref{fig13} shows that the external magnetic field is strong enough to enable a coupling to the plasma and Figure \ref{fig12} shows which part of the atmosphere, in regions of low magnetic dissipation,  can be transported with the magnetic field. For this, we calculate both the decoupled diffusion coefficient, $\eta_d$, and the Ohmic diffusion coefficients of the dominating ions identified in Section \ref{iv.ii}.
The individual Ohmic diffusion coefficients  provide insight in to where these ions are coupled to the potential magnetic field:
If $\eta_{ohm}$ is high, there is large magnetic dissipation, hence potential Ohmic heating.
Contrarily, if $\eta_{ohm}$ is low, the magnetic dissipation is small meaning a strong coupling of plasma to the magnetic field. If $\eta_{ohm}$ is low, the fluid motion of the plasma will transport a magnetic field.

At the substellar point (Fig.~\ref{fig12}, top left) almost all of the $\eta_{ohm}$'s (with $\eta_{ohm, K^+}$ being the exception) decrease with increasing atmospheric pressure on  HAT-P-7b, hence most of the atomic ions are coupled to a magnet field and would be transported along with the flow. $\eta_d$ and $\eta_{ohm,K^+}$, however, increases. This implies that, at the substellar point, the free electrons and $K^+$ become less strongly magnetically coupled and the other dominating electron donors become more coupled as we travel deeper in to HAT-P-7b's atmosphere.
At the antistellar point, the diffusion coefficients (Fig.~\ref{fig12}, top right)  increase deeper in to HAT-P-7b's atmosphere, except for the electrons and K$^+$. $\eta_d$ and $\eta_{ohm,K^+}$ decrease. This implies that, as the local gas pressure increases inward, free electrons and $K^+$ become more strongly coupled to a magnetic field whilst the other dominating electron donors become less so. 

Our results suggest that the entire atmosphere of HAT-P7b can be magnetised by a global magnetic field which may affect the global circulation of HAT-P-7b and  introduce atmospheric oscillation patterns as suggested by \cite{2017NatAs...1E.131R} with a typical time scale of $\sim$ 11.5 days.

\cite{2014ApJ...796...16K} investigate the electromagnetic behaviour of the atmosphere of the hot Jupiter HD\,209458~b and demonstrate that ionised hydrogen and helium (H$^+$, He$^+$) will sustain the ionosphere at p$_{\rm gas}<10^{-8}$ due to host-star photoionisation of the dayside.  A detailed analysis of conductivity regions is presented  and it is shown that the magnetic coupling in the atmosphere  increases with height for HD\,209458b.


\section{Detectability of HAT-P7b's ionosphere}\label{detectability}

Previous studies have shown that molecules in the upper dayside atmospheres of Ultra Hot-Jupiters (UHJs) should be mostly thermally dissociated at pressures probed by the transmission and emission spectroscopy at $p_{\rm gas}<10^{-3}$ bar. Therefore,  atomic neutrals and ions  are essential opacity contributors in the upper atmospheric layers of this class of planets \citep[e.g.][]{2018ApJ...866...27L,2019ApJ...876...69L,hoeijmakers2019spectral,yan2019ionized,hatp7b1}. Particularly, a recent study by \citet{2018ApJ...866...27L} suggests that Fe, Fe$^+$, Ti, Ni, Ca, Ca$^+$, SiO, and even relatively trace species such as Cr and Mn could contribute in the slopes and features shortward of 0.5~$\mu$m in transmission spectra of UHJs. 
Although these studies are based on low-resolution data, their conclusion should also hold true for high-resolution spectroscopy. In the present context, the major difference between low- and high-resolution spectroscopy is that the information on the continuum level would be lost in the high-res due to normalization of the spectra. But the spectral signatures of these species are expected to prevail.

Observation of the dominant ions (i.e. Fe$^+$, Al$^+$, Mg$^+$, Ca$^+$, Si$^+$, K$^+$, Na$^+$, Li$^+$) would provide a direct mean to detect the ionosphere of HAT-P-7b. In principle, these ions are accessible by high-resolution ground-based observations and through two methods: 1) detection of the resolved lines, and 2) cross correlation function (CCF) method. Sodium doublet lines at around 589.0 nm and 589.6 nm and Ca$^+$ H\&K (393.4 nm and 396.8 nm) and Ca$^+$ IRT (849.9 nm, 854.2 nm, and 866.2 nm) are among the most commonly detected resolved lines due to their large absorption cross section and high abundance \citep[e.g.][]{khalafinejad2017exoplanetary,casasayas2017detection,yan2019ionized}. The CCF method, on the other hand, utilises and combines the information of the lines’ positions (and to a lesser extend their shapes) to enhance 
detection probability from the data.
For instance, Fe and Fe$^+$ are among the robustly detected atoms on ultra-hot exoplanets, given their large number of atomic lines \citep[e.g.][]{hoeijmakers2019spectral,yan2020temperature,2020ApJ...894L..27P}. However, there are several challenges for the detection of ions on HAT-P-7b with these methods that are discussed below.

One of these challenges is the lack of line-list data for the most of ions. In order to calculate the opacity tables of the dominant ions, we adopt the transition tables and partition functions from \cite{Kurucz2018}. Among the mentioned ions, K$^+$, Na$^+$, and Li$^+$ lack in line-list data. Such incompleteness in the line-lists is one of the major reasons for the low number of ion detections on UHJs, where such ions are expected to exist \citep[e.g.][]{hatp7b1,hoeijmakers2019spectral}.

For the ions with available line-lists, the number of spectral lines are relatively low at low temperatures, e.g. less than 5,000 K. Therefore, the detection of ions on exoplanets has been limited to species with either strong resolved spectral features (such as Ca$^+$; \citealt{yan2019ionized}) or high number of spectral lines (such as Fe$^+$; \citealt{hoeijmakers2019spectral,2020arXiv200611308H}).

In a transmission spectrum, through which the terminators are being probed, a net blueshift could correspond to a dayside-to-nightside atmospheric asymmetry. This can be measured both through an investigation of the resolved lines and CCF method. On the other hand, a broadened resolved line might indicate a zonal-jet-dominated pattern for a co-rotating ionosphere. In contrast, emission spectroscopy probes the dayside, which is more sensitive to the conditions at the substellar point. As a result, emission spectroscopy is less relevant to distinguish between these two atmospheric patterns. Consequently, transmission spectroscopy seems to be a more appropriate approach. However, lower temperatures at the terminators cause a lower degree of ionisation
relative to the substellar point; see e.g. Fig.~\ref{fig10}. Therefore, this lower ion number densities result in less pronounced ionic spectral features in the transmission spectrum.

\begin{figure*}
\centering
  \includegraphics[width=180mm]{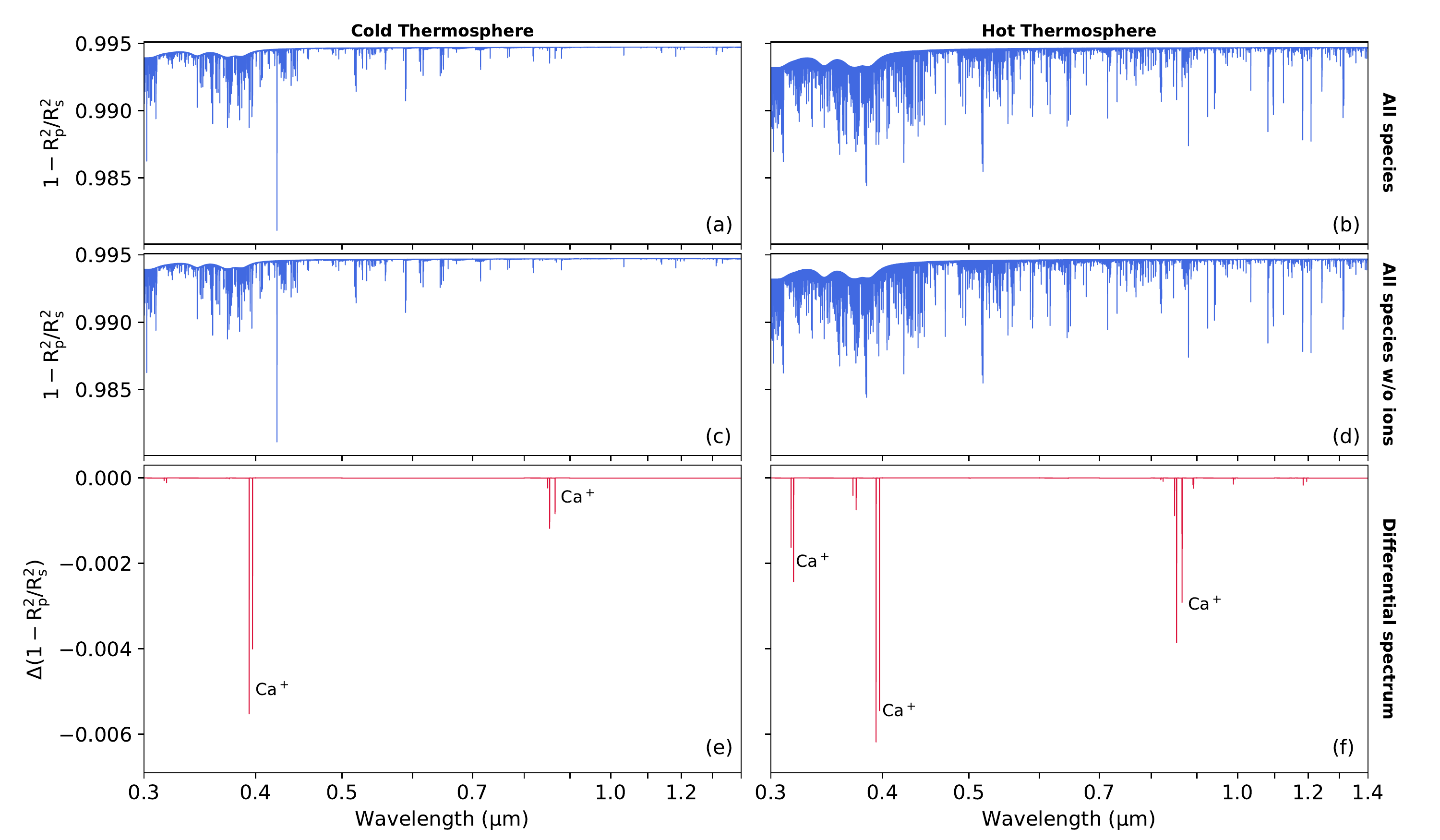}
\caption{\footnotesize (a) Synthetic high-resolution spectrum of HAT-P-7b over a wide wavelength range, assuming a cold thermosphere scenario with 5,000~K. (c) Similar to (a) except ion are excluded. (e) The difference between (a) and (c) that shows the significance of ion signatures. (b), (d), and (f) are similar to (a), (c), and (e) but assuming a hot thermosphere with 10,000~K. The pronounced ions are marked in the differential spectrum panels, (e) and (f).}\label{spectrum}
\end{figure*}

In addition to these difficulties, continuum opacities such as collision-induced absorption (CIA) of H$_2$-H$_2$ and H$_2$-He, continuum opacity due to a negatively-charged hydrogen ion (H$^-$), Rayleigh scattering due to  atomic and molecular hydrogen and helium, and clouds, obscure ionic spectral features. This is more significant at shorter wavelengths, where atomic and ionic spectral features are also expected to be more pronounced. Other sources of opacity, such as atomic (e.g. Fe) or molecular features, can also obscure ionic features, which will be discussed later.

To quantitatively assess the observability of ions on HAT-P-7b, we calculate  two synthetic high-resolution transmission spectra of this planet using petitRADTRANS \citep{molliere2019petitradtrans,molliere2020retrieving}.
 In both models, we assume the presence of atoms (Fe, Al, Si, Mg, Ca, Li, Na, and K), molecules (TiO, VO, CO, H$_2$O, H$_2$S, and SiO) and continuum opacities (CIA of H$_2$-H$_2$ and H$_2$-He, H$^-$, and Rayleigh scattering due to atomic and molecular hydrogen and helium), but in one model we exclude ions (Fe$^+$, Al$^+$, Mg$^+$, Ca$^+$, and Si$^+$). By taking the difference of these two spectra we estimate the significance of ionic features and the possibility of their observations.

Spectral line cores are usually sensitive to the pressures lower than 1~bar and mostly above mbar regime, but the upper boundary in GCMs simulations are usually extended up to 10$^{-3}$-10$^{-6}$~bar. In our GCM simulation, the upper boundary is set at around 1 $\mu$bar. Therefore, the radiative transfer model should be extended to lower pressures in order to properly calculate the high-resolution spectra. Consequently, we extend our radiative transfer models to 10$^{-12}$~bar. Current high-resolution spectrographs are not sensitive enough to allow for a complex temperature profile retrieval from their observations and hence we assume an isothermal condition for this extended part of the models.

As noted earlier, ions have limited line list information for temperatures below 5,000~K. As a result, we assume a thermospheric temperature of 5,000~K as the ``cold thermosphere'' scenario, where at or above this temperature access to an ionosphere through the observation of resolved lines might be more feasible. But observations suggest that an exoplanet atmosphere may extend into an upper thermosphere of exoplanets that is likely to be hotter \citep[e.g.][]{yan2020temperature}. A similar scenario has been proposed for Brown Dwarf - White Dwarf binaries in \cite{2017MNRAS.471.1728L}. We investigate such a scenario by setting the upper thermosphere temperature to 10,000~K, i.e. the ``hot thermosphere'' scenario.

The results are shown in Figure \ref{spectrum} where (a) illustrates synthetic high-resolution spectrum of HAT-P-7b between 0.3 and 1.4~$\mu$m, assuming a cold thermosphere scenario with 5,000~K. Ions are excluded from the spectrum shown in (c) and the differential spectrum of (a) and (c) are shown in (e). The right panels in Figure \ref{spectrum}, i.e. (b), (d), and (f), are similar to (a), (c), and (e) but for a hot thermosphere scenario. The pronounced ions in the differential spectra are labeled. Three outcomes immediately emerge: 1) the continuum obscures all ion features, 2) Ca$^+$ lines are the dominant observable ionic spectral features, while all other ions appear to contribute insignificantly (and hence their disappearance in the differential spectra), and 3) as expected, the spectral lines become more significant when the hot thermosphere is in place.

Our results suggests that the detection of ions on HAT-P-7b is most likely limited by the continuum. Therefore, Ca$^+$ strongest features (i.e. Ca$^+$ H\&K) at round 393.4 nm and 396.8 nm likely to be the only observable ionic features. While a hotter thermosphere scenario obscures ionic lines due to higher number of atomic spectral features, ionic spectral lines are similarly become pronounced. As a result, the relative significance of the ionic lines remain similar. This can be seen by comparing the panels (e) and (f) in Figure \ref{spectrum}. Nevertheless, Ca$^+$ H\&K remain the best candidates to investigate the presence of a co-rotating ionosphere on HAT-P-7b. Some atomic lines, such as Fe, Al, and Na, may also be detectable, but not useful as direct tracers of an ionosphere.

\section{Conclusions}\label{v}
We continued our investigation of the atmosphere of the ultra-hot Jupiter HAT-P-7b and present a study of the changing ionisation in the atmosphere of HAT-P-7b by means of calculating the plasma and magnetic properties of the atmosphere. Only thermal ionisation is considered in this study, the effects of irradiation from the host star or cosmic ray ionisation, for example, are not studied but will extend the ionosphere as well as the impact of  electromagnetic processes. We conclude that:

$-$ HAT-P-7b has a considerable degree of thermal ionisation at deep atmosphere levels across the whole globe. The degree of ionisation is considerably lower on nightside than on the dayside in the upper atmosphere.

$- $ An extended ionosphere exists on the dayside of HAT-P-7b which extends deep into the atmosphere. 

$-$ The relevant length scales effected by electromagnetic interactions in the gas-phase are larger in low density regions of HAT-P-7b across the whole planet.

$-$ Electromagnetic interactions dominate over electron-neutral interactions at all points in the atmosphere on the day side of HAT-P-7b. Higher up in the atmosphere on the nightside of the planet,  electron-neutral interactions may dominate over long-range, electromagnetic interactions of charged particles.

$-$ Fe$^+$,  Al$^+$, Si$^+$, Na$^+$, Mg$^+$, K$^+$, Ca$^+$ and Li$^+$ are the dominating ions in all areas of the atmosphere of HAT-P-7b, whilst K$^+$, Na$^+$, Li$^+$, Ca$^+$, Al$^+$ provide the majority of free electrons.

$-$ The Ca$^+$\,H\&K lines are the best candidates to investigate the presence of a co-rotating ionosphere on HAT-P-7b. Fe, Al and Na may also be detectable, but not as ionosphere tracer.

$-$ Cloud formation effects the abundances of important electron donors and have therefore an indirect impact on the ionisation of HAT-P-7b.

$-$  Discharge processes in form of lightning  may occur at  the morning terminator ($\phi=-90.0^{\circ}, \theta=0.0^{\circ}$) at on the nightside in the cloudy parts of the atmosphere of  HAT-P7b.

$-$ The minimum threshold for the magnetic flux density required for electrons to be magnetised is smaller than Jupiter's global magnetic field strengths.  This supports the suggestion by \cite{2017NatAs...1E.131R} that HAT-P-7b's dayside may exhibit oscillation pattern of the time scale of 11.5 days.

\begin{acknowledgements}
We thank the referee for valuable input which helped us to improve the paper. MW thanks the members The Exoplanet and Planet-Forming Discs Research Group for valuable discussions.
DS acknowledges financial support from the Science and Technology Facilities Council (STFC), UK, for his PhD studentship (project reference 2093954). ChH acknowledges funding from the European Union H2020-MSCA-ITN-2019 under the Grant Agreement no. 860470 (CHAMELEON).
\end{acknowledgements}

\bibliographystyle{aa}
\bibliography{Reportbib}

\end{document}